\documentclass[10pt, 3p, fleqn]{elsarticle}
\usepackage[british]{babel}
\usepackage[utf8]{inputenc}
\usepackage{graphics}
\usepackage{graphicx}
\usepackage{amsmath,amssymb,esint}
\usepackage{lineno}
\usepackage{multirow}
\usepackage{subfig}
\usepackage{natbib}
\usepackage{eucal}
\usepackage{mathtools}
\usepackage{placeins}
\usepackage{enumerate}
\usepackage[usenames,x11names]{xcolor}
\usepackage[colorlinks]{hyperref}
\usepackage{setspace}
\usepackage{MnSymbol}
\usepackage[british]{babel}
\usepackage{booktabs}
\usepackage{siunitx}
\usepackage{tikz}
\usetikzlibrary{calc}
\usepackage{pgfplots}
\usepackage{pgfplotstable}
\usetikzlibrary{shapes.geometric, arrows, intersections, through}
\usetikzlibrary{arrows.meta}
\tikzset{>={Latex[width=1mm,length=1mm]}}
\usepackage{comment}
\usepackage{graphicx}
\graphicspath{{../ManuscriptFigures/}}

\usepackage{filecontents}
\usepackage[ruled,vlined]{algorithm2e}
\usepackage{varwidth}
\usepackage{adjustbox}

\usetikzlibrary{calc}
\tikzset{math3d/.style={z={(-0.65cm,-0.30cm)},y={(0cm,1cm)},x={(0.9cm,-0.15cm)}}}
\usetikzlibrary{shapes.geometric, arrows, intersections, through}
\usetikzlibrary{decorations.text, decorations.shapes, backgrounds}
\usetikzlibrary{arrows.meta}
\usetikzlibrary{colorbrewer}
\usetikzlibrary{patterns}
\usepgfplotslibrary{colorbrewer}
\usepgfplotslibrary{colormaps}

\tikzstyle{process} = [rectangle, minimum width=4.1cm, minimum height=1cm, text centered, text width = 4.1cm, draw=black]
\tikzstyle{decision} = [diamond, minimum width=4.1cm, minimum height=1cm, aspect=2,inner sep=-0.5ex,text centered, text width = 4.1cm, draw=black]
\tikzstyle{arrow} = [thick,->,>=stealth]
\tikzstyle{line} = [draw, -latex']

\biboptions{sort&compress}

\begin{document}

\begin{frontmatter}

\title{Fully coupled implicit finite-volume algorithm for viscoelastic interfacial flows}

\author[affil2]{Ayman Mazloum}
\author[affil2]{Gabriele Gennari}
\author[affil3]{Fabian Denner}
\author[affil2]{Berend van Wachem\corref{cor1}}
\ead{berend.van.wachem@multiflow.org}

\address[affil2]{Chair of Mechanical Process Engineering, Otto-von-Guericke-Universit\"{a}t Magdeburg, Universit\"atsplatz 2, 39106 Magdeburg, Germany}
\address[affil3]{Department of Mechanical Engineering, Polytechnique Montréal, Montréal, H3T 1J4, Québec, Canada}

\cortext[cor1]{Corresponding author: }

\begin{abstract}
A fully coupled implicit finite-volume algorithm for incompressible viscoelastic interfacial flows is proposed, whereby the viscoelasticity of the flow is described by an upper-convected Maxwell constitutive model, including limited extensibility and shear-thinning behaviour. The governing equations describing the conservation of continuity and momentum, as well as the constitutive model are discretized using standard finite-volume methods and are solved for pressure, velocity and the polymer stress tensor in a single linear system of equations. Treating all terms of the linearized and discretized governing equations implicit in velocity, pressure and/or the components of the polymer stress tensor, a tightly coupled system of equations is obtained. The interface separating the interacting bulk phases and the surface tension acting at the fluid interface are modelled using a state-of-the-art front-tracking method. We demonstrate the capabilities of the proposed numerical framework with four representative test cases, including the deformation of a viscoelastic droplet in shear flow at large Weissenberg numbers of up to $\mathrm{Wi} = 10^4$, and the jump discontinuity of the rise velocity of a bubble rising in a viscoelastic liquid as a result of a ``negative wake''. Contrary to previous studies using segregated algorithms, the proposed fully coupled implicit algorithm does not apply or require a log-conformation approach to predict these flows. Overall, the fully implicit coupled front-tracking formulation provides a robust framework to reliable numerical predictions of strongly elastic interfacial flows at large Weissenberg numbers.
\end{abstract}
\end{frontmatter}

\section{Introduction}

Classical engineering applications of viscoelastic fluids, such as the lubrication dynamics in bearings \citep{Williamson1997,Gamaniel2021} and binders for composite materials \citep{Wei2019b}, as well as emerging manufacturing techniques by 3D printing \citep{Yuk2018, Nelson2020} and the assembly of soft materials \citep{Frumkin2021, Brun2022} have been driving an increased interest in viscoelastic interfacial flows, in which two or more immiscible fluids interact with each other, and the physical phenomena associated with these flows. The airborne dispersion of respiratory diseases, e.g.~corona viruses, gave the interest in viscoelastic interfacial flows a further boost in the wake of the recent pandemic, since mucus, the primary carrier of the viral load, exhibits dominant viscoelastic properties \citep{Lai2009}. As a result of these relevant and timely applications of viscoelastic interfacial flows, their numerical modelling has become a subject of growing activity in the scientific community. Even though computational rheology has developed into a mature discipline in recent decades, a large variety of constitutive models for the viscoelastic stress (typically referred to as the {\it polymer stress}) based on often competing assumptions \citep{Siginer2014}, large differences in numerical predictions and different interpretations of the underpinning physical phenomena \citep{Dupret1985,Alves2021}, as well as convergence difficulties for flows with even moderate elasticity, known as the high-Weissenberg number problem \citep{Keunings1986, Tsai2000, Walters2003}, are still hampering a widespread adoption and application of numerical modelling tools for viscoelastic flows. 

State-of-the-art algorithms for incompressible viscoelastic flows mostly rely on segregated algorithms \citep{Alves2021}, in which the momentum equations, a pressure projection equation satisfying the continuity constraint and the constitutive model describing the viscoelastic stresses are solved sequentially. However, the weak explicit coupling between velocity, pressure and polymer stress in the discretized governing equations as a result of the iterative predictor-corrector solution procedure severely limits the stability and convergence of these algorithms, requiring a strong underrelaxation of the discretized equations to reach a converged solution \citep{Keshtiban2004,Habla2013,Fernandes2019}. 
Numerical methods to mitigate problems associated with a high elasticity of the fluid, most notably the log-conformation approach \citep{Fattal2004, Fattal2005, Becker2023,Doherty2024}, have laid the foundation for new developments that substantially expanded the parameter range, especially with respect to the Weissenberg number, that can now be simulated routinely with widely available computational tools \citep{Hulsen2005,Afonso2009,Habla2013,Martins2015,Niethammer2018,Fernandes2022}, including viscoelastic multiphase flows \citep{Tome2012, Niethammer2019, Fernandes2019a, Naseer2023}.

Coupled implicit algorithms, whereby all governing equations are solved simultaneously in a single linear system of implicitly coupled equations, present a powerful alternative to the widely used segregated algorithms. This class of algorithms, which has been applied successfully to incompressible and compressible Newtonian single-phase \citep{Xiao2017, Denner2020, Darwish2009a,Darwish2014,Chen2010} and interfacial \citep{Denner2014a, Denner2018b, Denner2022c, Gorges2022, Darwish2015} flows, allows a tight implicit coupling of the governing equations, improving the stability and performance of the solution algorithm. \citet{Fernandes2019} presented a coupled implicit finite-volume algorithm for two-dimensional viscoelastic flows, whereby the constitutive model for the polymer stress tensor is included in the linear system of implicitly coupled equations and the polymer stress tensor is treated implicitly in the momentum equations. Shortly after, \citet{Pimenta2019} proposed a fully coupled algorithm for three-dimensional electrically driven viscoelastic flows using a Poisson-Nernst-Planck model, further adding a Poisson equation governing the electrical potential and transport equations for the charge densities to the linear system of implicitly coupled equations. 
Recently, \citet{Fernandes2022} presented a fully coupled algorithm for two-dimensional viscoelastic flows applying a log-conformation approach to the constitutive model.
All three studies \citep{Fernandes2022,Fernandes2019,Pimenta2019} reported a robust and rapid convergence with an impressive speed-up exceeding one order of magnitude for some cases 
compared to a state-of-the-art segregated algorithm for viscoelastic flows. 
While the simulation of Newtonian interfacial flows, in which two (or more) immiscible fluids interact with each other, has become common and a large variety of numerical frameworks based on the volume-of-fluid \citep{Hirt1981}, level-set \citep{Osher1988,Sussman1994} and front-tracking \citep{Unverdi1992,Tryggvason2001} methods are available, simulation tools that can accurately predict three-dimensional interfacial flows in which at least one of the interacting fluids is viscoelastic are still scarce. Finite-volume algorithms for viscoelastic interfacial flows using volume-of-fluid \citep{Habla2011, Niethammer2019, Giliberto2026}, level-set \citep{Pillapakkam2001,Kabanemi2020,Amani2020} and front-tracking \citep{Izbassarov2015,Naseer2023} methods have previously been presented, alongside algorithms based on the Lattice-Boltzmann method \citep{Wang2022b} and diffusive interface method \citep{Yue2004,Yue2006,Rodriguez2019,Zografos2020}. For incompressible viscoelastic interfacial flows, state-of-the-art algorithms rely on solving for the conformation tensor of the polymer stress rather than for the polymer stress tensor directly, and current finite-volume algorithms are built upon segregated algorithms. However, a fully coupled implicit algorithm for viscoelastic interfacial flows has not yet been presented in the literature.
In the context of Newtonian flows, fully coupled implicit algorithms have been demonstrated to be particularly suited for interfacial flows as they can be readily applied to interfacial flows with large density ratios. Additionally, the capillary time-step constraint, a severe impediment for the simulation of flows with surface tension \citep{Denner2015, Popinet2018}, can be breached when surface tension is treated implicitly \citep{Denner2022b}. 


Building upon recently published algorithms on viscoelastic single-phase flows \citep{Fernandes2019, Pimenta2019, Fernandes2022} and our own prior work on fully coupled implicit algorithms for Newtonian flows \citep{Denner2014a, Denner2020, Denner2022b, Janodet2025}, we propose a fully coupled implicit finite-volume algorithm for incompressible viscoelastic interfacial flows. To account for the viscoelasticity of the flow, we consider the upper-convected Maxwell constitutive equation for the polymer stress tensor, including limited extensibility and shear-thinning behaviour, demonstrated with the linear and exponential Phan-Thien-Tanner models \citep{Phan-Thien1977}, as well as the Giesekus model \citep{Giesekus1982}. The governing equations describing the conservation of continuity and momentum, as well as the constitutive model are discretized using standard finite-volume methods and are solved for pressure, velocity and the polymer stress tensor in a single linear system of equations. The interface separating the interacting bulk phases and the surface tension acting at the fluid interface are modelled using a state-of-the-art front-tracking method \citep{Gorges2022, Gorges2023}. We demonstrate the capabilities of the proposed numerical framework with representative test cases, including the deformation of a viscoelastic droplet in shear flow at large Weissenberg numbers and the jump discontinuity of the rise velocity of a bubble rising in a viscoelastic liquid. Notably, the proposed algorithm does not apply nor require a log-conformation approach to predict these flows, contrary to the studies presented in the literature to date.

Section \ref{sec:governing} introduces the governing conservation laws and the considered constitutive model. The discretization of these governing equations and the numerical framework are described in Section \ref{sec:numerics}, where we examine the discretization of each term of the governing equations, and the front-tracking method employed to represent and transport the fluid interface is briefly reviewed in Section \ref{sec:ft}. A presentation of the complete discretized governing equations and an explanation of the solution procedure is the subject of Section \ref{sec:solution}. The results of four representative test cases are presented and discussed in Section \ref{sec:results}, and the article is concluded in Section \ref{sec:conclusion}.

\section{Governing equations}
\label{sec:governing}

The considered incompressible and isothermal viscoelastic flows are governed by the continuity and momentum equations, given as
\begin{align}
  \nabla \cdot \mathbf{u} &= 0 \label{eq:continuity} \\
  \rho \left[ \frac{\partial \mathbf{u}}{\partial t} +  \nabla \cdot (\mathbf{u} \otimes \mathbf{u}) \right] &= \nabla \cdot \boldsymbol{\varsigma} + \mathbf{S}_\sigma + \mathbf{S}_g, \label{eq:momentum}
\end{align}
respectively, where $t$ denotes time, $\mathbf{u}\equiv (u~v~w)^\text{T}$ denotes the velocity vector, $\rho$ is the fluid density, $\mathbf{S}_\sigma$ is the source term representing surface tension (when an interface is present) and $\mathbf{S}_g=\rho \, \mathbf{g}$ is the source term representing gravity, with $\mathbf{g}$ the gravitational acceleration. The stress tensor
$\boldsymbol{\varsigma} = -  p \, \mathbf{I} + \boldsymbol{\tau}_\mathrm{s} + \boldsymbol{\tau}_\mathrm{p}$ comprises the normal stress tensor exerted by pressure, $-p \, \mathbf{I}$, where $\mathbf{I}$ denotes the identity tensor, the solvent stress tensor $\boldsymbol{\tau}_\mathrm{s}$ and the polymer stress tensor $\boldsymbol{\tau}_\mathrm{p}$, such that the momentum equations follow as
\begin{equation}
  \rho \left[ \frac{\partial \mathbf{u}}{\partial t} +  \nabla \cdot (\mathbf{u} \otimes \mathbf{u}) \right] = - \nabla p + \nabla \cdot  \boldsymbol{\tau}_\mathrm{s} + \nabla \cdot \boldsymbol{\tau}_\mathrm{p} + \mathbf{S}_\sigma + \mathbf{S}_g.
\end{equation}
The solvent stress tesnor is based on Newton's law of viscosity and given as $\boldsymbol{\tau}_\mathrm{s} =2 \mu \mathbf{D}$,   
where $\mu$ is the dynamic solvent viscosity, and $\mathbf{D} =\frac{1}{2} (\nabla \mathbf{u} + \nabla \mathbf{u}^\mathrm{T})$ is the rate of deformation tensor.

The non-Newtonian viscoelastic behaviour of the flow is expressed by the evolution of the polymer stress tensor $\boldsymbol{\tau}_\mathrm{p}$, 
which, in turn, is governed by a general differential constitutive equation of the form \citep{Alves2021}
\begin{equation}
  \psi \, \boldsymbol{\tau}_\mathrm{p} + \lambda \left(\frac{\partial \boldsymbol{\tau}_{\mathrm{p}}}{\partial t} + \mathbf{u} \cdot \nabla \boldsymbol{\tau}_\mathrm{p} - \boldsymbol{\tau}_\mathrm{p} \cdot \nabla \mathbf{u} - \nabla \mathbf{u}^\mathrm{T} \cdot \boldsymbol{\tau}_\mathrm{p} + \xi\left(\boldsymbol{\tau}_p\mathbin{\cdot}\mathbf{D}
+\mathbf{D}\mathbin{\cdot}\boldsymbol{\tau}_p\right) \right)  + \frac{\alpha \, \lambda}{\eta} \, \boldsymbol{\tau}_\mathrm{p} \cdot \boldsymbol{\tau}_\mathrm{p}  = 2 \eta \mathbf{D} , \label{eq:constitutive}
\end{equation}
where $\psi$ is the stress function, $\lambda$ is the relaxation time, $\xi$ is the non-affine or ``slip'' parameter, $\alpha$ is the mobility parameter and $\eta$ is the polymer viscosity. The first four terms gathered in big parentheses in Eq.~\eqref{eq:constitutive} together constitute the upper-convected time derivative of the polymer stress tensor, where the third and fourth terms ensure the correct transformation of $\boldsymbol{\tau}_\mathrm{p}$ under deformations by the flow \citep{Snoeijer2020}, and the fifth term captures the slip between the molecular network and the continuum medium \citep{Alves2001}. 
Note that the polymer stress tensor is symmetric and, therefore, has only 6 unique components.
As constitutive models we consider 
the Giesekus model \citep{Giesekus1982}, for which $0 < \alpha \leq 0.5$, $\psi=1$ and $\xi=0$, the 
linear Phan-Thien-Tanner model (LPTT) \citep{Phan-Thien1977}, for which $\alpha=0$ and 
$\psi = 1 + \lambda \, \epsilon \, \text{tr}(\boldsymbol{\tau}_{\mathrm{p}})/\eta$,
and the exponential Phan-Thien-Tanner model (EPTT) \citep{Phan-Thien1977}, for which $\alpha=0$ and 
$\psi = \exp \left[\lambda \, \epsilon \, \text{tr}(\boldsymbol{\tau}_{\mathrm{p}})/\eta \right]$,
where $\epsilon$ is the extensibility coefficient of the fluid. 
All three models are widely used in the literature and convey the primary features of modelling viscoelastic flows.

In order to distinguish the interacting fluid phases, we consider a generic indicator function $\mathcal{I}$ that is reconstructed based on the position of the interface. The indicator function is defined as
\begin{equation}
  \mathcal{I}(\mathbf{x}) = \begin{cases}
    0, &\mbox{if} \ \mathbf{x} \in \Omega_\mathrm{a}\\
    1, &\mbox{if} \ \mathbf{x} \in \Omega_\mathrm{b}
  \end{cases}
\end{equation} 
where $\Omega = \Omega_\text{a} \cup \Omega_\text{b}$ is the computational domain, with $\Omega_\text{a}$ and $\Omega_\text{b}$ the subdomains occupied by fluids ``a'' and ``b'', respectively. Following previous work on viscoelastic interfacial flows \citep{Izbassarov2015,Rodriguez2019}, the fluid properties $\phi \in \{\alpha,\eta,\lambda,\mu,\rho,\psi,\,\xi \}$ are defined based on the indicator function as $\phi(\mathbf{x}) = \phi_\mathrm{a} + \mathcal{I}(\mathbf{x}) (\phi_\mathrm{b}-\phi_\mathrm{a})$.
Surface tension is modelled as a volumetric source term $\mathbf{S}_\sigma$ in the momentum equations,
\begin{equation}
  \mathbf{S}_\sigma = \sigma \kappa \mathbf{n}_\mathrm{\Sigma} \delta_\mathrm{\Sigma},
\end{equation} 
where $\sigma$ is the surface tension coefficient, $\kappa$ is the interface curvature, $\mathbf{n}_\Sigma$ is the normal vector of the interface and $\delta_\Sigma$ is the interfacial delta function.

\section{Numerical framework}
\label{sec:numerics}
The proposed numerical framework is implemented in our in-house finite-volume solver \texttt{MultiFlow} and is built upon a collocated second-order finite-volume discretization and a fully coupled implicit solution algorithm \citep{Denner2020, Bartholomew2018}, whereby the discretized governing equations are solved in a single system of linearized equations, $\mathbf{A} \cdot \boldsymbol{\zeta} = \mathbf{b}$,
with the three velocity components, pressure and the six unique components of the polymer stress tensor as the implicitly sought solution variables.
Below we describe the discretization of the individual terms of the governing conservation laws, Eqs.~\eqref{eq:continuity} and \eqref{eq:momentum}, and the constitutive model, Eq.~\eqref{eq:constitutive}, as well as the stress-velocity coupling.
The interface is modelled using a front-tracking method \citep{Gorges2022}, presented in Section \ref{sec:ft}, although any other suitable method to capture or track the interface, e.g.,~volume-of-fluid or level-set methods, may equally be applied together with the proposed numerical framework.

\subsection{Discretization methods}

The discretization is based on a standard second-order finite-volume method. Discretizing the generic convection-diffusion equation of a general fluid variable $\phi$,
\begin{equation}
  \frac{\partial \phi}{\partial t} + \nabla \cdot (\mathbf{u} \phi) = \nabla \cdot (D_\phi \, \nabla \phi) + S,
\end{equation}
where $D_\phi$ is the diffusion coefficient of $\phi$ and $S$ is a generic source term, with the employed second-order finite-volume method is given in semi-discretized form for cell $P$ of an arbitrary computational mesh as
\begin{equation}
  \left. \frac{\partial \phi}{\partial t} \right|_P  V_P +  \sum_f \tilde{\phi}_f \, F_f = \sum_f D_{\phi,f} \, (\nabla \phi_f \cdot \mathbf{n}_f) \, A_f +  S_P \, V_P,
\end{equation}
where $f$ denotes all faces bounding mesh cell $P$, $\tilde{\square}$ denotes a flux-limited interpolation, $\mathbf{n}_f$ is the normal vector of face $f$ pointing out of cell $P$, and the area of face $f$ and the volume of cell $P$ are denoted with $A_f$ and $V_P$, respectively. 
The flux $F_f$ through face $f$ is defined as
\begin{equation}
  F_f = \vartheta_f A_f, \label{eq:flux_simple}
\end{equation}
where the advecting velocity $\vartheta_f=\mathbf{u}_f \cdot \mathbf{n}_f$ is obtained by a momentum-weighted interpolation (MWI), as discussed in detail in Section \ref{sec:mwi}.
The transient derivative is discretized using the second-order backward Euler (BDF2) scheme for a variable time-step as \citep{Moukalled2016}
\begin{equation}
  \left. \frac{\partial \phi}{\partial t} \right|_P  \approx \left(\frac{1}{\Delta t_1} + \frac{1}{\Delta t_1 + \Delta t_2} \right) \phi_P^{(t)} - \left(\frac{1}{\Delta t_1} + \frac{1}{\Delta t_2} \right) \phi_P^{(t-\Delta t_1)} + \frac{\Delta t_1}{\Delta t_1 \Delta t_2 + \Delta t_2^2} \, \phi_P^{(t-\Delta t_1-\Delta t_2)}, \label{eq:bdf2}
\end{equation}
where $\Delta t_1$ is the current time-step and $\Delta t_2$ is the previous time-step. The superscripts $(t)$, $(t-\Delta t_1)$ and $(t-\Delta t_1-\Delta t_2)$ denote the solution at the current time-level, the previous time-level, and the previous-previous time-level, respectively.
In the discretization of the advection term,  $\phi_f$ is interpolated from the values at adjacent cell centers using a flux-limited interpolation scheme, given as 
\begin{equation}
  \tilde{\phi}_f = \phi_U + \chi_f (\phi_D - \phi_U), \label{eq:fluxlimited_interpolation}
\end{equation}
where $\chi_f$ denotes the flux limiter, and subscripts $D$ and $U$ denote the downwind cell and the upwind cell of face $f$, respectively. In this study, we consider the central-differencing scheme as well as the CUBISTA scheme \citep{Alves2003}, which is widely used to compute viscoelastic flows, to determine the flux limiter $\chi_f$, but other suitable schemes may equally be applied.
The face-centered velocity gradient projected along the normal vector of the cell face, $\nabla \phi_f \cdot \mathbf{n}_f$, in the discretized diffusion term is decomposed into an orthogonal and a non-orthogonal part to correct for any non-orthogonality of the computational mesh, following the work of \citet{Demirdzic1995}, as
\begin{equation}
  \nabla \phi_f \cdot \mathbf{n}_f \approx c_f \, \frac{\phi_Q - \phi_P}{\Delta s_f} + \overline{\nabla \phi}_f \cdot (\mathbf{n}_f - c_f \, \mathbf{s}_f), \label{eq:nonorthogonalcorr}
\end{equation}
where $c_f = (\mathbf{n}_f \cdot \mathbf{s}_f)^{-1}$ is the scaling factor of the decomposition \citep{Mathur1997} and where $\overline{\square}_f = (1-\ell_f) \square_P + \ell_f \square_Q$ denotes a linear interpolation, with $\ell_f$ the inverse-distance weighting coefficient with respect to cell $P$ and face $f$. The vector $\mathbf{s}_f$ is the unit vector connecting the cell centers adjacent to face $f$, pointing from cell $P$ to neighbour cell $Q$ and  $\Delta s_f = |\mathbf{s}_f|$ is the distance between the centers of cells $P$ and $Q$. 




\subsection{Continuity equation}
\label{sec:continuity}

Applying the divergence theorem, the continuity equation, Eq.~\eqref{eq:continuity}, is readily discretized using the flux $F_f$ through face $f$, as defined in Eq.~\eqref{eq:flux_simple}, as
\begin{equation}
  \iiint_{V_P} \nabla \cdot \mathbf{u} \, \text{d}V \approx \sum_f F_f^{(n+1)}  = 0. \label{eq:continuity_disc}
\end{equation}
To this end, the implicitly treated advecting velocity given by the MWI, see Section \ref{sec:mwi}, of the form
\begin{equation}
  \vartheta_f^{(n+1)} = \overline{\mathbf{u}}_f^{(n+1)} \cdot \mathbf{n}_f + f \Big(\nabla p^{(n+1)}, \mathbf{u}^{(t-\Delta t_1)}\Big)
\end{equation}
introduces an implicit dependency of the continuity equation on pressure, such that the continuity equation can be solved implicitly for velocity and pressure in a fully coupled manner \citep{Denner2020}.

The iteration counter $n$ is associated with nonlinear iterations performed to solve the system of discretized governing equations at each time-step, as further explained in Section \ref{sec:solution}, with superscript $(n)$ denoting deferred quantities and superscript $(n+1)$ denoting quantities for which the solution is sought implicitly.

\subsection{Momentum equations}
\label{sec:numerics_momentum}

The momentum equations are solved implicitly in velocity, pressure and the components of the polymer stress tensor.
The transient term is discretized, at mesh cell $P$, as
\begin{equation}
  \iiint_{V_P} \rho \, \frac{\partial \mathbf{u}}{\partial t} \ \text{d}V \approx  \rho_P \left. \frac{\partial \mathbf{u}}{\partial t} \right|_P^{(n+1)} V_P \label{eq:momentum_transient}
\end{equation}
with the transient derivative of velocity following from Eq.~\eqref{eq:bdf2} as
\begin{equation}
  \left. \frac{\partial \mathbf{u}}{\partial t} \right|_P^{(n+1)} = \left(\frac{1}{\Delta t_1} + \frac{1}{\Delta t_1 + \Delta t_2} \right) \mathbf{u}_P^{(n+1)} - \left(\frac{1}{\Delta t_1} + \frac{1}{\Delta t_2} \right) \mathbf{u}_P^{(t-\Delta t_1)} + \frac{\Delta t_1}{\Delta t_1 \Delta t_2 + \Delta t_2^2} \, \mathbf{u}_P^{(t-\Delta t_1-\Delta t_2)}.
\end{equation}
The advection term is discretized by applying a Newton linearization treating both the flow velocity $\mathbf{u}$ and the flux $F$ implicitly,
\begin{equation}
  \iiint_{V_P}  \rho \, \nabla \cdot (\mathbf{u} \otimes \mathbf{u}) \ \text{d}V \approx  \rho_P \, \sum_f \left(\tilde{\mathbf{u}}_f^{(n+1)} \, F_f^{(n)} + \tilde{\mathbf{u}}_f^{(n)} \,F_f^{(n+1)} - \tilde{\mathbf{u}}_f^{(n)} \, F_f^{(n)}\right), \label{eq:momentum_advection_newton}
 \end{equation}
where $\tilde{\square}$ denotes the flux-limited interpolation presented in Eq.~\eqref{eq:fluxlimited_interpolation}.

The divergence of the stress tensor appearing on the right-hand side of the momentum equations, Eq.~\eqref{eq:momentum}, is split into three parts, $\nabla \cdot \boldsymbol{\varsigma} = -  \nabla p + \nabla \cdot \boldsymbol{\tau}_\mathrm{s} + \nabla \cdot \boldsymbol{\tau}_\mathrm{p}$, each of which is discretized separately. The pressure gradient, $\nabla p = \nabla \cdot (p\mathbf{I})$, and the divergence of the polymer stress tensor, $\nabla \cdot \boldsymbol{\tau}_\mathrm{p}$, are both discretized using linear interpolation of the respective cell values to the mesh faces as
\begin{align}
  &\iiint_{V_P} \nabla p \ \text{d}V \approx  \sum_f \overline{p}_f^{(n+1)} \, \mathbf{n}_f A_f \label{eq:momentum_pressuregrad} \\
  &\iiint_{V_P} \nabla \cdot {\boldsymbol{\tau}}_{\mathrm{p}} \ \text{d}V \approx \sum_f \left(\overline{\boldsymbol{\tau}}_{\mathrm{p},f}^{(n+1)} \cdot \mathbf{n}_f\right)  A_f. \label{eq:momentum_nablataup}
\end{align}
Both $p$ and ${\boldsymbol{\tau}}_{\mathrm{p}}$ are treated implicitly, as indicated by the superscript $(n+1)$. The divergence of the solvent stress tensor $\boldsymbol{\tau}_{\mathrm{s}}$ is discretized as \citep{Denner2020}
\begin{equation}
 \iiint_{V_P} \nabla \cdot \boldsymbol{\tau}_{\mathrm{s}} \ \text{d}V \approx \sum_f  \breve{\mu}_f \left(\nabla \mathbf{u}_f \cdot \mathbf{n}_f +  \overline{\nabla \mathbf{u}}_f^{\mathrm{T}} \cdot \mathbf{n}_f \right) A_f. \label{eq:momentum_shear_general}
\end{equation}
where, including the correction for non-orthogonal meshes presented in Eq.~\eqref{eq:nonorthogonalcorr},  
\begin{equation}
  \nabla \mathbf{u}_f \cdot \mathbf{n}_f \approx c_f \, \frac{\mathbf{u}_Q - \mathbf{u}_P}{\Delta s_f} + \overline{\nabla \mathbf{u}}_f \cdot (\mathbf{n}_f - c_f \mathbf{s}_f). 
\end{equation}
A harmonic interpolation is applied to interpolate the viscosity values from the cell centers $P$ and $Q$ to the shared face $f$ \citep{Ferziger2003}, 
\begin{equation}
  \frac{1}{\breve{\mu}_f} = \frac{1-\ell_f}{\mu_P} + \frac{\ell_f}{\mu_Q},
\end{equation}
where $\breve{\mu}_f$ is the harmonic interpolated face-centered viscosity, and $\ell_f$ is the inverse-distance weighting coefficient with respect to cell $P$ and face $f$.
The solvent stress term is implemented treating all velocity terms implicitly, such that
\begin{equation}
  \iiint_{V_P} \nabla \cdot \boldsymbol{\tau}_{\mathrm{s}} \ \text{d}V \approx \sum_f  \mu \left[c_f \, \frac{\mathbf{u}_Q^{(n+1)} - \mathbf{u}_P^{(n+1)}}{\Delta s_f} + \overline{\nabla \mathbf{u}}_f^{(n+1)} \cdot (\mathbf{n}_f - c_f \mathbf{s}_f) +  \overline{\nabla \mathbf{u}}_f^{\mathrm{T},(n+1)} \cdot \mathbf{n}_f \right] A_f. \label{eq:momentum_shear_all}
\end{equation} 
Contrary to previous studies on fully coupled algorithms for Newtonian flows, e.g.~\citep{Denner2020,Darwish2009a}, and viscoelastic flows \citep{Fernandes2019}, all velocity gradients of the discretized solvent shear stress in Eq.~\eqref{eq:momentum_shear_all} are solved implicitly.

As shown in previous studies \citep{Mencinger2007, Denner2014a, Bartholomew2018}, the momentum sources $\mathbf{S}$ have to be discretized with the same discretization as the pressure gradients, for the discretized pressure gradient $\nabla p_P$ to be able to match the discretized source term $\mathbf{S}^\star_P$.
Following \citet{Bartholomew2018}, the discretized momentum source $\mathbf{S}^\star$ is constructed based on the untreated source term $\mathbf{S}$ as
\begin{equation}
  \mathbf{S}^\star_P = \frac{1}{V_P} \sum_f \left( \overline{\mathbf{S}}_f \cdot \mathbf{s}_f \right)  \Delta s_f \mathbf{n}_f A_f.
\end{equation}

\subsection{Constitutive model}
\label{sec:numerics_constitutive}

The constitutive model, Eq.~\eqref{eq:constitutive}, yields six governing equations for the six unique components of the polymer stress tensor, which are solved implicitly for the components of the polymer stress tensor, as well as velocity and pressure.

In the constitutive model, the first term on the left-hand side of Eq.~\eqref{eq:constitutive} is treated implicitly as
\begin{equation}
  \iiint_{V_P} \psi \, \boldsymbol{\tau}_\text{p} \ \mathrm{d}V \approx \psi_P \, \boldsymbol{\tau}_{\text{p},P} ^{(n+1)} \, V_P
  \label{eq:constitutive_stressfunction}
\end{equation}
and the transient term of the constitutive model, as part of the upper-convected derivative of the polymer stress tensor, is discretized in the same manner as transient term of the momentum equations, 
\begin{equation}
  \iiint_{V_P} \lambda  \frac{\partial \boldsymbol{\tau}_\text{p}}{\partial t} \ \mathrm{d}V \approx \lambda_P \left. \frac{\partial \boldsymbol{\tau}_\text{p}}{\partial t} \right|_P^{(n+1)}  V_P,  \label{eq:constitutive_transient}
\end{equation}
using the second-order backward Euler scheme presented in Eq.~\eqref{eq:bdf2}.

The advection term of the constitutive model, Eq.~\eqref{eq:constitutive}, arises from the material derivative of $\boldsymbol{\tau}_\text{p}$ and is, consequently, not in conserved form, contrary to the advection term of the momentum equations. In the interest of a discretization that is consistent with the advection of momentum, we reformulate the advection term of $\boldsymbol{\tau}_\text{p}$ using the product rule as
\begin{equation}
  \mathbf{u} \cdot \nabla \boldsymbol{\tau}_\mathrm{p} = \nabla  \cdot (\mathbf{u} \cdot \boldsymbol{\tau}_\mathrm{p}) -  \boldsymbol{\tau}_\mathrm{p} \, (\nabla \cdot \mathbf{u}) , \label{eq:constitutive_advection_general}
\end{equation}
such that the advecting velocity can now be applied to define the fluxes through the mesh faces. Similar to the advection term, 
a Newton linearization is applied, with
\begin{equation}
  \iiint_{V_P} \lambda \, (\mathbf{u} \cdot \nabla \boldsymbol{\tau}_\mathrm{p}) \ \text{d}V \approx
  \lambda_P \left[ \sum_f \left( \tilde{\boldsymbol{\tau}}_{\mathrm{p},f}^{(n+1)} -\boldsymbol{\tau}_{\mathrm{p},P}^{(n+1)} \right) F_f^{(n)} + \sum_f \left( \tilde{\boldsymbol{\tau}}_{\mathrm{p},f}^{(n)} - \boldsymbol{\tau}_{\mathrm{p},P}^{(n)}\right) F_f^{(n+1)} - \sum_f \left(\tilde{\boldsymbol{\tau}}_{\mathrm{p},f}^{(n)} - \boldsymbol{\tau}_{\mathrm{p},P}^{(n)}\right) F_f^{(n)} \right] .
  \label{eq:constitutive_advection_newton}
\end{equation}
For the considered incompressible flows, the ($\nabla \cdot \mathbf{u}$)-term in Eq.~\eqref{eq:constitutive_advection_general} is superfluous from a mathematical viewpoint. Numerically, however, this is only true for a converged result, but may not be the case during the initial nonlinear iterations in each time-step. Including the $(\nabla \cdot \mathbf{u})$-term, therefore, generally improves the convergence of the solution algorithm.


Aiming to fully exploit the implicit coupling provided by the fully coupled solution procedure, the two remaining terms of the upper-convected time derivative of the polymer stress tensor and the non-affine response term are linearized and treated implicit using a Newton linearization respectively, 
\begin{multline}
  \iiint_{V_P} \lambda \left(\boldsymbol{\tau}_\mathrm{p} \cdot \nabla \mathbf{u} + \nabla \mathbf{u}^\mathrm{T} \cdot \boldsymbol{\tau}_\mathrm{p} \right) \ \text{d}V \approx \lambda_P \, \Big(
  \boldsymbol{\tau}_{\mathrm{p},P}^{(n+1)} \cdot \nabla \mathbf{u}^{(n)}_P
  + \boldsymbol{\tau}_{\mathrm{p},P}^{(n)} \cdot \nabla \mathbf{u}^{(n+1)}_P 
  - \boldsymbol{\tau}_{\mathrm{p},P}^{(n)} \cdot \nabla \mathbf{u}^{(n)}_P  \\ 
  + \nabla \mathbf{u}^{\mathrm{T},(n)}_P \cdot {\boldsymbol{\tau}_{\mathrm{p},P}^{(n+1)}} 
  + \nabla \mathbf{u}^{\mathrm{T},(n+1)}_P \cdot {\boldsymbol{\tau}_{\mathrm{p},P}^{(n)}} 
  - \nabla \mathbf{u}^{\mathrm{T},(n)}_P \cdot {\boldsymbol{\tau}_{\mathrm{p},P}^{(n)}} 
  \Big) \, V_P,
  \label{eq:constitutive_deformation_impl}
\end{multline}
\begin{equation}
\begin{aligned}
\iiint_{V_P} \lambda\,\xi\,(\boldsymbol{\tau}\!\cdot\!\mathbf{D}+\mathbf{D}\!\cdot\!\boldsymbol{\tau})\,\mathrm{d}V
\;\approx\;
\frac{\lambda_P\xi_P\,V_P}{2}\,\Big(
&\boldsymbol{\tau}_P^{(n+1)}\!\cdot\!\nabla\mathbf{u}_P^{(n)}
+\boldsymbol{\tau}_P^{(n)}\!\cdot\!\nabla\mathbf{u}_P^{(n+1)}
-\boldsymbol{\tau}_P^{(n)}\!\cdot\!\nabla\mathbf{u}_P^{(n)}
\\
&+\nabla\mathbf{u}_P^{(n)}\!\cdot\!\boldsymbol{\tau}_P^{(n+1)}
+\nabla\mathbf{u}_P^{(n+1)}\!\cdot\!\boldsymbol{\tau}_P^{(n)}
-\nabla\mathbf{u}_P^{(n)}\!\cdot\!\boldsymbol{\tau}_P^{(n)}
\\
&+\boldsymbol{\tau}_P^{(n+1)}\!\cdot\!\nabla\mathbf{u}_P^{\mathsf T,(n)}
+\boldsymbol{\tau}_P^{(n)}\!\cdot\!\nabla\mathbf{u}_P^{\mathsf T,(n+1)}
-\boldsymbol{\tau}_P^{(n)}\!\cdot\!\nabla\mathbf{u}_P^{\mathsf T,(n)}
\\
&+\nabla\mathbf{u}_P^{\mathsf T,(n)}\!\cdot\!\boldsymbol{\tau}_P^{(n+1)}
+\nabla\mathbf{u}_P^{\mathsf T,(n+1)}\!\cdot\!\boldsymbol{\tau}_P^{(n)}
-\nabla\mathbf{u}_P^{\mathsf T,(n)}\!\cdot\!\boldsymbol{\tau}_P^{(n)}
\Big),
\end{aligned}
\end{equation}
the quadratic stress term of the Giesekus model is treated implicitly with respect to the polymer stress tensor,
\begin{equation}
  \iiint_{V_P} \frac{\alpha \, \lambda}{\eta} \, \boldsymbol{\tau}_\mathrm{p} \cdot \boldsymbol{\tau}_\mathrm{p} \ \text{d}V \approx  \frac{\alpha_P \, \lambda_P}{\eta_P} \, \boldsymbol{\tau}_{\mathrm{p},P}^{(n)} \cdot \boldsymbol{\tau}_{\mathrm{p},P}^{(n+1)} \, V_P,
\end{equation}
and the strain-rate tensor is treated implicitly with respect to the velocity,
\begin{equation}
  \iiint_{V_P} \eta \left(\nabla \mathbf{u} + \nabla \mathbf{u}^\mathrm{T} \right) \ \text{d}V \approx \eta_P \left(\nabla \mathbf{u}_P^{(n+1)} + \nabla \mathbf{u}_P^{\mathrm{T},(n+1)} \right) V_P. 
  \label{eq:constitutive_strainrate_impl}
\end{equation}
\citet{Fernandes2019} treated both the additional terms of the upper-convective time derivative, Eq.~\eqref{eq:constitutive_deformation_impl}, and the strain-rate tensor, Eq.~\eqref{eq:constitutive_strainrate_impl}, explicitly. Recently, \citet{Fernandes2022} also applied a Newton linearization to the additional terms of the upper-convective time derivative in their fully coupled log-conformation algorithm, treating these terms implicitly in the conformation tensor and the velocity. \citet{Pimenta2019} treated the first term of Eq.~\eqref{eq:constitutive_strainrate_impl} implicitly with respect to velocity, while treating the second term explicitly.

Contrary to the fully coupled algorithm of \citet{Fernandes2019}, we do not divide the discretized constitutive model by the relaxation time $\lambda$ before discretization. Hence, the constitutive model is valid for $\lambda \geq 0$. Considering, for example, the upper-convected Maxwell model ($\psi=1$, $\mu=0$), Eq.~\eqref{eq:constitutive} reduces to $\boldsymbol{\tau}_\text{p} = \eta (\nabla \mathbf{u} + \nabla \mathbf{u}^\mathrm{T} )$ for $\lambda =0$, resulting in a  Newtonian flow with shear viscosity $\eta$.

\subsection{Stress-velocity coupling}
\label{sec:stressvelocitycoupling}

\begin{figure}
  \centering
  \subfloat[Small-stencil treatment of face $f$]{\includegraphics[scale=1]{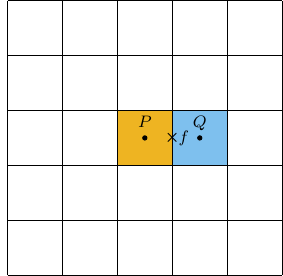}} \qquad
  \subfloat[Large-stencil treatment of face $f$]{\includegraphics[scale=1]{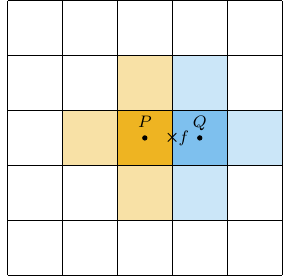}}
  \caption{Interpolation stencils of the velocity at face $f$, with the adjacent cells $P$ and $Q$, considered for the stress-velocity coupling.}
  \label{fig:stencils}
\end{figure}

The stress-velocity coupling associated with the polymer stress is a central building block of the proposed methodology. 
The commonly applied method to ensure a robust coupling between the polymer stress and the velocity when a collocated variable arrangement is used, is to introduce two mathematically equivalent diffusion terms with opposite signs on the right-hand side of the momentum equations, Eq.~\eqref{eq:momentum}, to yield
\begin{equation}
  \rho \left[ \frac{\partial \mathbf{u}}{\partial t} + \nabla \cdot (\mathbf{u} \otimes \mathbf{u}) \right] = - \nabla p + \nabla \cdot \boldsymbol{\tau}_\mathrm{s} + \nabla \cdot \boldsymbol{\tau}_\mathrm{p} + \mathbf{S}_\sigma + \mathbf{S}_g + \underbrace{\nabla \cdot (\hat{\eta} \, \nabla \mathbf{u})}_\text{small stencil} - \underbrace{\nabla \cdot (\hat{\eta} \, \nabla \mathbf{u})}_\text{large stencil}, \label{eq:momentum_bsd}
\end{equation}
where $\hat{\eta}$ is a weighting factor that is dimensionally equivalent to a dynamic viscosity.
As the notation indicates, the additional terms are discretized on different computational stencils. For clarity, we first consider an equidistant Cartesian mesh with $c_f=1$ and $\Delta s_f = \Delta x$, with the discretization for general non-orthogonal meshes given thereafter in Eqs.~\eqref{eq:bsd_small_disc} and \eqref{eq:bsd_large_disc}. On an equidistant Cartesian mesh, the \textit{small-stencil} diffusion term is discretized as
\begin{equation}
  \iiint_{V_P} \nabla \cdot (\hat{\eta} \, \nabla \mathbf{u})\ \mathrm{d}V \approx \sum_f \hat{\eta}  \frac{\mathbf{u}_Q-\mathbf{u}_P}{\Delta x}  A_f, \label{eq:bsd_small}
 \end{equation}
and the \textit{large-stencil} diffusion term as
\begin{equation}
 \iiint_{V_P} \nabla \cdot (\hat{\eta} \, \nabla \mathbf{u})\ \mathrm{d}V \approx \sum_f \hat{\eta} \left(\overline{\nabla \mathbf{u}}_f \cdot \mathbf{n}_f\right) A_f, \label{eq:bsd_large}
\end{equation}
as illustrated in Figure \ref{fig:stencils}.
In the literature, this procedure of adding two diffusion terms with opposite signs is widely referred to as \textit{both-sides diffusion} (BSD) \citep{Guenette1995}, typically applied with $\hat\eta=\eta$ \citep{Fernandes2019, Fernandes2022,Pimenta2017, Habla2013}. Applying the conventionally used centered finite-difference approximations described above, the additional small-stencil and large-stencil diffusion terms yield, using tensor notation and the Einstein summation convention,
\begin{equation}
 \hat{\eta} \left. \frac{\partial u_i}{\partial x_j} \right|_f - \frac{1}{2} \left(\hat{\eta} \left. \frac{\partial u_i}{\partial x_j} \right|_P + \hat{\eta} \left. \frac{\partial u_i}{\partial x_j} \right|_Q \right) \approx \hat{\eta} \, \frac{u_{i,W} - 3 u_{i,P} + 3 u_{i,E} - u_{i,EE}}{4 \Delta x_j} = - \hat{\eta} \left. \frac{\partial^3 u_i}{\partial x_j^3} \right|_f \frac{\Delta x_j^2}{4}, \label{eq:velocity_filter}
\end{equation}
which infers that, by taking the divergence of this term in the momentum equations, the two additional diffusion terms introduce numerical diffusion ($\partial u_i^4/\partial x_j^4$) with a magnitude proportional to $\hat{\eta}$ and $\Delta x^2$ \citep{Pimenta2017}.

In line with the discretization of the divergence of the solvent stress described in Eq.~\eqref{eq:momentum_shear_all} and with $\hat{\eta}=\breve{\eta}_f$, the small-stencil and large-stencil terms are discretized as
\begin{align}
  \iiint_{V_P} \nabla \cdot (\hat{\eta} \, \nabla \mathbf{u})\ \mathrm{d}V &\approx \sum_f \breve{\eta}_f \left[c_f \frac{\mathbf{u}_Q^{(n+1)}-\mathbf{u}_P^{(n+1)}}{\Delta s_f} + \overline{\nabla \mathbf{u}}_f^{(n+1)} \cdot (\mathbf{n}_f - c_f \mathbf{s}_f) \right] A_f \label{eq:bsd_small_disc} \\
 \iiint_{V_P} \nabla \cdot (\hat{\eta} \, \nabla \mathbf{u})\ \mathrm{d}V &\approx \sum_f \breve{\eta}_f \left(\overline{\nabla \mathbf{u}}_f^{(n+1)} \cdot \mathbf{n}_f\right) A_f, \label{eq:bsd_large_disc}
\end{align}
respectively. The contribution of the stress-velocity coupling to the right-hand side of the discretized momentum equations can, thus, be summarized as
\begin{align}
  \iiint_{V_P} [\underbrace{\nabla \cdot (\hat{\eta} \, \nabla \mathbf{u})}_{\text{small stencil}} - \underbrace{\nabla \cdot (\hat{\eta} \, \nabla \mathbf{u})}_{\text{large stencil}} ] \ \mathrm{d}V \approx \sum_f \breve{\eta}_f c_f  \left[\frac{\mathbf{u}_Q^{(n+1)}-\mathbf{u}_P^{(n+1)}}{\Delta s_f} - \overline{\nabla \mathbf{u}}_f^{(n+1)} \cdot \mathbf{s}_f\right] A_f. \label{eq:bsd_mf3_disc} 
\end{align}
To fully exploit the implicit coupling afforded by the fully coupled algorithm and to be consistent with the treatment of the strain-rate term in the constitutive model, all velocity contributions in Eq.~\eqref{eq:bsd_mf3_disc} are treated implicitly.

\subsection{Pressure-velocity coupling}
\label{sec:mwi}

To ensure a robust pressure-velocity coupling using the employed collocated variable arrangement, the flux $F_f = \vartheta_f A_f$ through face $f$ is defined with an advecting velocity $\vartheta_f = \mathbf{u}_f \cdot \mathbf{n}_f$ that is evaluated using a momentum-weighted interpolation (MWI) \cite{Bartholomew2018}, originally introduced by \citet{Rhie1983}. This advecting velocity allows to solve the continuity equation for pressure \citep{Denner2020} and prevents pressure-velocity decoupling on the employed collocated variable arrangement \citep{Bartholomew2018}.

The advecting velocity is, based on the unified formulation of the momentum-weighted interpolation proposed by \citet{Bartholomew2018}, defined as
\begin{equation}
  \vartheta_f = \overline{\mathbf{u}}_{f} \cdot \mathbf{n}_{f} - \hat{d}_f \left[ \left(  \nabla p_f -  \overline{\nabla p}_f \right) \cdot \mathbf{s}_{f} - \left(  \mathbf{S}_f -  \overline{\mathbf{S}}^\star_f \right) \cdot \mathbf{s}_{f} -  \frac{\breve{\rho}_f}{\Delta t_1} \left( \vartheta_f^{(t-\Delta t_1)} - \overline{\mathbf{u}}_{f}^{(t-\Delta t_1)} \cdot \mathbf{n}_{f} \right) \right],
  \label{eq:mwi}
\end{equation}
where $\breve{\rho}_f$
is the harmonically averaged face density.
The weighting factor $\hat{d}_f$ defines the strength of the pressure-velocity coupling and is given as
\begin{equation}
  \hat{d}_f = \dfrac{ \dfrac{V_P}{a_P} + \dfrac{V_Q}{a_Q}}{2+ \dfrac{\breve{\rho}_f}{\Delta t_1} \left(\dfrac{V_P}{a_P} + \dfrac{V_Q}{a_Q}\right)}. \label{eq:dhat}
\end{equation}
The coefficients $a_P$ and $a_Q$ are defined based on the diagonal matrix coefficients of the velocity arising from the advection term, see Eq.~\eqref{eq:momentum_advection_newton}, the solvent stress term, see Eq.~\eqref{eq:momentum_shear_all}, and the small-stencil stress-coupling term, see Eq.~\eqref{eq:bsd_mf3_disc}, of the discretized momentum equations associated with the cells adjacent to face $f$. For the discretization presented above, the coefficient $a_P$ (and, analogously, $a_Q$) is given as 
\begin{equation}
  a_P = \rho_P \displaystyle\sum_f F_f \chi_f^\prime + \displaystyle\sum_f \left(\breve{\mu}_f + \breve{\eta}_f \right) \dfrac{c_f  A_f}{\Delta s_f},
\end{equation}
where $\chi_f^\prime = (1-\chi_f)$ if $F_f \geq 0$ and $\chi_f^\prime = \chi_f$ if $F_f < 0$.
For the MWI to be time-step independent, $a_P$ and $a_Q$ must not include the contribution of the transient terms to the diagonal coefficient \citep{Bartholomew2018}.

For an arbitrary unstructured mesh and including a density-weighting of the large-stencil pressure and source term contributions, the discretized and implicitly treated advecting velocity is defined as \citep{Bartholomew2018}
\begin{multline}
  \vartheta_f^{(n+1)} = \overline{\mathbf{u}}_{f}^{(n+1)} \cdot \mathbf{n}_{f} - \hat{d}_f \left[ \frac{p_Q^{(n+1)}-p_P^{(n+1)}}{\Delta s_f} - \frac{\breve{\rho}_f}{2} \left( \frac{\nabla p_P^{(n+1)}}{\rho_P} + \frac{\nabla p_Q^{(n+1)}}{\rho_Q} \right) \cdot \mathbf{s}_f   \right. \\ 
  - \left. \mathbf{S}_f^{(n)} \cdot  \mathbf{s}_f + \frac{\breve{\rho}_f}{2} \left( \frac{\mathbf{S}_P^{\star,(n)}}{\rho_P} + \frac{\mathbf{S}_Q^{\star,(n)}}{\rho_Q} \right) \cdot \mathbf{s}_f -  \frac{\breve{\rho}_f}{\Delta t_1} \left( \vartheta_f^{(t-\Delta t_1)} - \overline{\mathbf{u}}_{f}^{(t-\Delta t_1)} \cdot \mathbf{n}_{f} \right)\right].
  \label{eq:mwi_disc}
\end{multline}
The discretized pressure terms together constitute a low-pass filter on the pressure field that prevents pressure-velocity decoupling \citep{Bartholomew2018}.
Contrary to most previous work on fully coupled algorithms \citep{Denner2020, Darwish2009a,Fernandes2019}, we treat all pressure terms in Eq.~\eqref{eq:mwi_disc} implicitly \citep{Janodet2025}.

\section{Front-tracking method}
\label{sec:ft}

The numerical framework proposed in the previous section is complemented by a front-tracking method \citep{Gorges2022} to track the fluid interface separating to immiscible bulk phases. Since the precise formulation and implementation of the applied interface tracking (or interface capturing) method is not critical for the proposed numerical framework, we only provide a brief overview of the applied front-tracking method and refer the reader to our recent publications \citep{Gorges2022, Gorges2023} for more details.

In front tracking \citep{Unverdi1992, Tryggvason2001}, the fluid interface is represented by a triangulated surface mesh.
Each vertex $i$ of this surface mesh is advected in a Lagrangian manner,
\begin{equation}
  \frac{\mathrm{d}\mathbf{x}_i(t)}{\mathrm{d}t} = \mathbf{u}(\mathbf{x}_i,t), \label{eq:dxdt_ft}
\end{equation}
where $\mathbf{x}_i$ and ${\mathbf{u}}$ are the location and (interpolated) velocity of vertex $i$, respectively. The vertices of the surface mesh can, consequently, also move tangential to the interface, which may lead to vertex clustering and a deteriorating quality of the surface mesh, in turn requiring extensive remeshing of the surface mesh to retain an acceptable mesh quality. In order to address this issue, we apply the {\it normal-only advection} (NOA) of the vertices \citep{Gorges2023}, with the velocity at the location of the surface mesh vertices defined as 
\begin{equation}
  \mathbf{u}(\mathbf{x}_i,t) = \mathbf{u}_\mathrm{ref}(t) + \left\{\left[\overline{\mathbf{u}}(\mathbf{x}_i,t) - \mathbf{u}_\mathrm{ref}(t)\right] \cdot \mathbf{n}(\mathbf{x}_i,t) \right\} \mathbf{n}(\mathbf{x}_i,t)
\end{equation}
where $\mathbf{u}_\mathrm{ref}(t)$ is a spatially invariant reference velocity and $\overline{\mathbf{u}}(\mathbf{x}_i,t) $ is the interpolated fluid velocity at the location of vertex $i$. Since the fluid velocity is only known at the cell centers of the fluid mesh, the velocity $\overline{\mathbf{u}}(\mathbf{x}_i,t)$ at the location of the vertices of the surface mesh is interpolated from the fluid mesh using a Peskin cosine interpolation kernel \citep{Peskin1977}, 
\begin{equation}
  \overline{\mathbf{u}}(\mathbf{x}_i,t) = \sum_{L} \left[\mathbf{u}_L \, \prod_{j=1}^3 d\left(\frac{x_{j,i} - x_{j,L}}{\Delta x}\right)\right] ,
\end{equation}
where $L$ denotes all mesh cells in a $2 \Delta x \times 2 \Delta x \times 2 \Delta x$ stencil with respect to vertex $i$, and the weighting kernel is
\begin{equation}
  d(r) = \begin{cases}
    \dfrac{1}{4} \left[ 1+ \cos \left( \dfrac{\pi}{2} \, r \right) \right] , & \mbox{if} \ |r|<2\\
     0 ,& \mbox{if} \ |r|\geq 2.
  \end{cases}
\end{equation}
We define the reference velocity as the volume-averaged velocity of the body enclosed by the front,
\begin{equation}
  \mathbf{u}_\mathrm{ref} \approx \frac{\sum_P \mathbf{u}_P \, \mathrm{I}_\mathrm{P} \, V_P}{\sum_P  \mathrm{I}_\mathrm{P} \, V_P},
\end{equation}
where $P$ denotes all cells of the fluid mesh, and integrate Eq.~\eqref{eq:dxdt_ft} using a conventional fourth-order Runge-Kutta scheme \citep{Gorges2023}.

The indicator function $\mathcal{I}$ is reconstructed based on the location of the surface mesh by solving a Poisson equation \citep{Tryggvason2011a}. The force due to surface tension is computed at each triangle $T$ of the surface mesh using a Frenet-Element algorithm \citep{Tryggvason2011a}
\begin{equation}
  \mathbf{F}_{\sigma,T} = \sigma \iint_{A_T} \kappa \mathbf{n} \, \mathrm{d}A = \sigma \oint_{l_e} \mathbf{p} \, \mathrm{d}l = \sigma \sum_e \mathbf{p}_e l_e,
\end{equation}
where $e$ denotes the edges of triangle $T$ with length $l_e$, outward-pointing planar vector $\mathbf{p}_e = \mathbf{n}_e \times \mathbf{t}_e$, normal vector $\mathbf{n}_e$ and tangential vector $\mathbf{t}_e$.
Subsequently, the force due to surface tension computed on the surface mesh is interpolated to the fluid mesh using the Peskin cosine interpolation kernel to define the surface tension source term as
\begin{equation}
  \mathbf{S}_{\sigma,P}  = \sum_T \left[\mathbf{F}_{\sigma,T} \, \prod_{j=1}^3 d\left(\frac{{x}_{j,P}-{x}_{j,T}}{\Delta x} \right)\right]. 
\end{equation}
where $T$ are all surface triangles in a $2 \Delta x \times 2 \Delta x \times 2 \Delta x$ stencil with respect to cell $P$.

The surface mesh is dynamically adapted to ensure a sufficient mesh quality as well as an adequate resolution of the interface, including a parabolic fit vertex repositioning method that reduces shape errors of the interface, as described in detail in our previous work \citep{Gorges2022}. In addition, we apply a volume correction step \citep{Pivello2012} to improve volume conservation and treat small undulations of the surface mesh in areas where the interface strongly contracts with the TSUR3D algorithm \citep{deSousa2004}. 


\section{Solution procedure}
\label{sec:solution}

Combining the discretization of the individual terms presented in Section \ref{sec:numerics}, we obtain a set of discretized equations governing the considered incompressible and isothermal viscoelastic flows.
The discretized continuity equation is given by Eq.~\eqref{eq:continuity_disc},
the discretized momentum equations 
are
\begin{multline}
  \rho_P \left[ \left. \frac{\partial \mathbf{u}}{\partial t} \right|_P^{(n+1)} V_P 
  +  \sum_f \left(\tilde{\mathbf{u}}_f^{(n+1)} \, F_f^{(n)} + \tilde{\mathbf{u}}_f^{(n)} \, F_f^{(n+1)} - \tilde{\mathbf{u}}_f^{(n)} \, F_f^{(n)} \right) \right] = - \sum_f \overline{p}_f^{(n+1)} \mathbf{n}_f A_f  \\ + \sum_f  \breve{\mu}_f \left[c_f \, \frac{\mathbf{u}_Q^{(n+1)}-\mathbf{u}_P^{(n+1)}}{\Delta s_f} + \overline{\nabla \mathbf{u}}_f^{(n+1)} \cdot (\mathbf{n}_f - c_f \, \mathbf{s}_f) +  \overline{\nabla \mathbf{u}}_f^{\mathrm{T}, (n+1)} \cdot \mathbf{n}_f \right] A_f + \sum_f \left(\overline{\boldsymbol{\tau}}_{\mathrm{p},f}^{(n+1)} \cdot \mathbf{n}_f\right) A_f \\
  + \mathbf{S}_{\sigma,P}^{\star,(n)} V_P + \mathbf{S}_{g,P}^{\star,(n)} V_P + \sum_f \breve{\eta}_f c_f \left[\frac{\mathbf{u}_Q^{(n+1)}-\mathbf{u}_P^{(n+1)}}{\Delta s_f} - \overline{\nabla \mathbf{u}}_f^{(n+1)} \cdot  \mathbf{s}_f  \right] A_f, \label{eq:momentum_disc_final}
\end{multline}
and the discretized constitutive model is given as
\begin{multline}
  \psi_P \, \boldsymbol{\tau}_{\mathrm{p},P}^{(n+1)} V_P + \lambda_P \Bigg[ \! \! \left. \frac{\partial \boldsymbol{\tau}_{\mathrm{p}}}{\partial t}\right|_P^{(n+1)} \! \! V_P + \sum_f \left( \tilde{\boldsymbol{\tau}}_{\mathrm{p},f}^{(n+1)} -\boldsymbol{\tau}_{\mathrm{p},P}^{(n+1)} \right) F_f^{(n)} + \sum_f \left( \tilde{\boldsymbol{\tau}}_{\mathrm{p},f}^{(n)} - \boldsymbol{\tau}_{\mathrm{p},P}^{(n)}\right) F_f^{(n+1)} - \sum_f \left(\tilde{\boldsymbol{\tau}}_{\mathrm{p},f}^{(n)} - \boldsymbol{\tau}_{\mathrm{p},P}^{(n)}\right) F_f^{(n)}  \\
  - \left(\boldsymbol{\tau}_{\mathrm{p},P}^{(n+1)} \cdot \nabla \mathbf{u}^{(n)}_P
  + \boldsymbol{\tau}_{\mathrm{p},P}^{(n)} \cdot \nabla \mathbf{u}^{(n+1)}_P 
  - \boldsymbol{\tau}_{\mathrm{p},P}^{(n)} \cdot \nabla \mathbf{u}^{(n)}_P    
  + \nabla \mathbf{u}^{\mathrm{T},(n)}_P \cdot {\boldsymbol{\tau}_{\mathrm{p},P}^{(n+1)}} 
  + \nabla \mathbf{u}^{\mathrm{T},(n+1)}_P \cdot {\boldsymbol{\tau}_{\mathrm{p},P}^{(n)}} 
  - \nabla \mathbf{u}^{\mathrm{T},(n)}_P \cdot {\boldsymbol{\tau}_{\mathrm{p},P}^{(n)}} 
  \right) V_P \\
  + \frac{\xi_P}{2}\,\Bigl(\boldsymbol{\tau}_P^{(n+1)}\!\cdot\!\nabla\mathbf{u}_P^{(n)}
+\boldsymbol{\tau}_P^{(n)}\!\cdot\!\nabla\mathbf{u}_P^{(n+1)}
-\boldsymbol{\tau}_P^{(n)}\!\cdot\!\nabla\mathbf{u}_P^{(n)}
+\nabla\mathbf{u}_P^{(n)}\!\cdot\!\boldsymbol{\tau}_P^{(n+1)}
+\nabla\mathbf{u}_P^{(n+1)}\!\cdot\!\boldsymbol{\tau}_P^{(n)}
-\nabla\mathbf{u}_P^{(n)}\!\cdot\!\boldsymbol{\tau}_P^{(n)} 
+\boldsymbol{\tau}_P^{(n+1)}\!\cdot\!\nabla\mathbf{u}_P^{\mathsf T,(n)}\\
+\boldsymbol{\tau}_P^{(n)}\!\cdot\!\nabla\mathbf{u}_P^{\mathsf T,(n+1)}
-\boldsymbol{\tau}_P^{(n)}\!\cdot\!\nabla\mathbf{u}_P^{\mathsf T,(n)}
+\nabla\mathbf{u}_P^{\mathsf T,(n)}\!\cdot\!\boldsymbol{\tau}_P^{(n+1)}
+\nabla\mathbf{u}_P^{\mathsf T,(n+1)}\!\cdot\!\boldsymbol{\tau}_P^{(n)}
-\nabla\mathbf{u}_P^{\mathsf T,(n)}\!\cdot\!\boldsymbol{\tau}_P^{(n)}
\Bigr)\,V_P
+ \frac{\alpha_P}{\eta_P} \, \boldsymbol{\tau}_{\mathrm{p},P}^{(n)} \cdot \boldsymbol{\tau}_{\mathrm{p},P}^{(n+1)} \, V_P \Bigg] \\ = \eta_P \left(\nabla \mathbf{u}_P^{(n+1)} + \nabla \mathbf{u}_P^{\mathrm{T},(n+1)} \right) V_P. \label{eq:constitutive_disc_final}
\end{multline}
As the notation suggests, each term of the governing equations makes an implicit contribution to at least one of the solution variables $\Gamma = \{p,u,v,w,\tau_{\text{p},xx},\tau_{\text{p},yy}, \tau_{\text{p},zz}, \tau_{\text{p},xy}, \tau_{\text{p},xz}, \tau_{\text{p},yz}\}$. 


\begin{figure}[t]
  \begin{center}
  \begin{small}
      \begin{tikzpicture}
      \node (pro0) [process] {Update previous time-levels: \\[0.25em] $\Gamma^{(t-\Delta t_1-\Delta t_2)}\leftarrow \Gamma^{(t-\Delta t_1)}$ \\[0.25em] $\Gamma^{(t-\Delta t_1)}\leftarrow \Gamma^{(n+1)\phantom{...)}}$  \\[0.25em] $F_f^{(t-\Delta t_1)} \leftarrow F_f^{(n+1)\phantom{....}}$};
      \node (pro0b) [process, below of=pro0, yshift=-1cm] {Advect the interface and compute the surface tension}; 
      \node (pro1a) [process, below of=pro0b, yshift=-0.5cm] {\mbox{Gather the coefficients} and assemble \mbox{$\mathbf{A}$ and $\mathbf{b}$}}; 
      \node (pro1b) [process, below of=pro1a, yshift=-0.5cm] {Solve $\mathbf{A} \cdot \boldsymbol{\zeta} = \mathbf{b}$}; 
      \node (pro3) [process, below of=pro1b, yshift=-0.5cm] {Compute $F_f^{(n+1)}$};
      \node (dec1) [decision, below of=pro3, yshift=-1cm]
      {Conservation satisfied?}; 
      \node (pro4) [process, below of=dec1, yshift=-1.2cm] {\mbox{Adapt the fluid mesh,} if applicable}; 
      \draw [arrow] (pro0) -- (pro0b);
      \draw [arrow] (pro0b) -- (pro1a);
      \draw [arrow] (pro1a) -- (pro1b);
      \draw [arrow] (pro1b) -- (pro3);
      \draw [arrow] (pro3) -- (dec1);
      \draw [arrow] (dec1) --+(-3cm,0) |- (pro1a);
      \draw [arrow] (dec1) -- (pro4);
      \draw [arrow] (pro4) --+(+3cm,0) |- (pro0);
      \node at (-2.5,-8.35) {no};
      \node at (0.35,-9.8) {yes};
      \node [rotate=90] at (-3.2,-6.2) {$n \leftarrow n+1$};
      \node [rotate=90] at (3.2,-6.2) {$t \leftarrow t+\Delta t$};
      \end{tikzpicture} 
  \end{small}
  \caption{Flow chart of the solution procedure of the discretized and linearized system of governing equations, where $n$ is the nonlinear iteration counter, $\Gamma = \{u,v,w,p,\tau_{\text{p},xx},\tau_{\text{p},yy}, \tau_{\text{p},zz}, \tau_{\text{p},xy}, \tau_{\text{p},xz}, \tau_{\text{p},yz}\}$ are the solution variables and $F_f$ is the flux through mesh face $f$ (see Section \ref{sec:mwi}). The coefficient matrix $\mathbf{A}$ holds all coefficients for the implicitly sought solution variables $\Gamma^{(n+1)}$ of the discretized governing equations and $\boldsymbol{\zeta}$ is the solution vector. The right-hand side vector $\mathbf{b}$ holds the deferred contributions of the previous iteration ($\Gamma^{(n)}$, $\vartheta_f^{(n)}$) and the contributions of the previous time-levels ($\Gamma^{(t-\Delta t_1)}$, $\Gamma^{(t-\Delta t_1-\Delta t_2)}$, $\vartheta_f^{(t-\Delta t_1)}$).}
\label{fig:flowchart}
\end{center}
\end{figure}
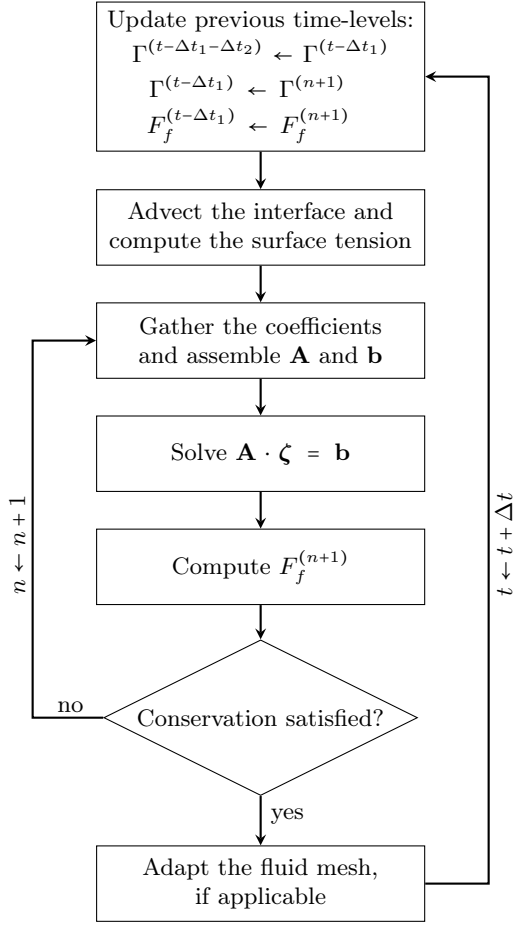

The solution procedure applied to solve the discretized governing equations is illustrated in Figure \ref{fig:flowchart}. In each time-step, the interface is advected first and, subsequently, the linearized and discretized governing equations \eqref{eq:continuity_disc}, \eqref{eq:momentum_disc_final} and \eqref{eq:constitutive_disc_final} are solved simultaneously in a single linear system of equations. 
For a three-dimensional computational mesh with $N$ cells, this linear system of equations is given as
\begin{equation}
  \begin{pmatrix}
    \boldsymbol{\mathcal{A}}_p & 
    \boldsymbol{\mathcal{A}}_u & 
    \boldsymbol{\mathcal{A}}_v & 
    \boldsymbol{\mathcal{A}}_w & 
    \mathbf{0} & 
    \mathbf{0} & 
    \mathbf{0} & 
    \mathbf{0} & 
    \mathbf{0} & 
    \mathbf{0} 
    \\[0.2em]
    \boldsymbol{\mathcal{B}}_p & 
    \boldsymbol{\mathcal{B}}_u & 
    \boldsymbol{\mathcal{B}}_v & 
    \boldsymbol{\mathcal{B}}_w & 
    \boldsymbol{\mathcal{B}}_{\tau_{\text{p},xx}} & 
    \mathbf{0} & 
    \mathbf{0} & 
    \boldsymbol{\mathcal{B}}_{\tau_{\text{p},xy}} & 
    \boldsymbol{\mathcal{B}}_{\tau_{\text{p},xz}} & 
    \mathbf{0} 
    \\[0.2em]
    \boldsymbol{\mathcal{C}}_p & 
    \boldsymbol{\mathcal{C}}_u & 
    \boldsymbol{\mathcal{C}}_v & 
    \boldsymbol{\mathcal{C}}_w & 
    \mathbf{0} & 
    \boldsymbol{\mathcal{C}}_{\tau_{\text{p},yy}} & 
    \mathbf{0} & 
    \boldsymbol{\mathcal{C}}_{\tau_{\text{p},xy}} & 
    \mathbf{0} & 
    \boldsymbol{\mathcal{C}}_{\tau_{\text{p},yz}}
    \\[0.2em]
    \boldsymbol{\mathcal{D}}_p & 
    \boldsymbol{\mathcal{D}}_u & 
    \boldsymbol{\mathcal{D}}_v & 
    \boldsymbol{\mathcal{D}}_w & 
    \mathbf{0} & 
    \mathbf{0} & 
    \boldsymbol{\mathcal{D}}_{\tau_{\text{p},zz}} & 
    \mathbf{0} & 
    \boldsymbol{\mathcal{D}}_{\tau_{\text{p},xz}} & 
    \boldsymbol{\mathcal{D}}_{\tau_{\text{p},yz}}
    \\[0.2em]
    \boldsymbol{\mathcal{E}}_p & 
    \boldsymbol{\mathcal{E}}_u & 
    \boldsymbol{\mathcal{E}}_v & 
    \boldsymbol{\mathcal{E}}_w & 
    \boldsymbol{\mathcal{E}}_{\tau_{\text{p},xx}} & 
    \mathbf{0} & 
    \mathbf{0} & 
    \boldsymbol{\mathcal{E}}_{\tau_{\text{p},xy}} & 
    \boldsymbol{\mathcal{E}}_{\tau_{\text{p},xz}} & 
    \mathbf{0} 
    \\[0.2em]
    \boldsymbol{\mathcal{F}}_p & 
    \boldsymbol{\mathcal{F}}_u & 
    \boldsymbol{\mathcal{F}}_v & 
    \boldsymbol{\mathcal{F}}_w & 
    \mathbf{0} & 
    \boldsymbol{\mathcal{F}}_{\tau_{\text{p},yy}} & 
    \mathbf{0} & 
    \boldsymbol{\mathcal{F}}_{\tau_{\text{p},xy}} & 
    \mathbf{0} & 
    \boldsymbol{\mathcal{F}}_{\tau_{\text{p},yz}}
    \\[0.2em]
    \boldsymbol{\mathcal{G}}_p & 
    \boldsymbol{\mathcal{G}}_u & 
    \boldsymbol{\mathcal{G}}_v & 
    \boldsymbol{\mathcal{G}}_w & 
    \mathbf{0} & 
    \mathbf{0} & 
    \boldsymbol{\mathcal{G}}_{\tau_{\text{p},zz}} & 
    \mathbf{0} & 
    \boldsymbol{\mathcal{G}}_{\tau_{\text{p},xz}} & 
    \boldsymbol{\mathcal{G}}_{\tau_{\text{p},yz}}
    \\[0.2em]
    \boldsymbol{\mathcal{H}}_p & 
    \boldsymbol{\mathcal{H}}_u & 
    \boldsymbol{\mathcal{H}}_v & 
    \boldsymbol{\mathcal{H}}_w & 
    \boldsymbol{\mathcal{H}}_{\tau_{\text{p},xx}} & 
    \boldsymbol{\mathcal{H}}_{\tau_{\text{p},yy}} & 
    \mathbf{0} & 
    \boldsymbol{\mathcal{H}}_{\tau_{\text{p},xy}} & 
    \boldsymbol{\mathcal{H}}_{\tau_{\text{p},xz}} & 
    \boldsymbol{\mathcal{H}}_{\tau_{\text{p},yz}}
    \\[0.2em]
    \boldsymbol{\mathcal{I}}_p & 
    \boldsymbol{\mathcal{I}}_u & 
    \boldsymbol{\mathcal{I}}_v & 
    \boldsymbol{\mathcal{I}}_w & 
    \boldsymbol{\mathcal{I}}_{\tau_{\text{p},xx}} & 
    \mathbf{0} & 
    \boldsymbol{\mathcal{I}}_{\tau_{\text{p},zz}} & 
    \boldsymbol{\mathcal{I}}_{\tau_{\text{p},xy}} & 
    \boldsymbol{\mathcal{I}}_{\tau_{\text{p},xz}} & 
    \boldsymbol{\mathcal{I}}_{\tau_{\text{p},yz}}
    \\[0.2em]
    \boldsymbol{\mathcal{J}}_p & 
    \boldsymbol{\mathcal{J}}_u & 
    \boldsymbol{\mathcal{J}}_v & 
    \boldsymbol{\mathcal{J}}_w & 
    \mathbf{0} & 
    \boldsymbol{\mathcal{J}}_{\tau_{\text{p},yy}} & 
    \boldsymbol{\mathcal{J}}_{\tau_{\text{p},zz}} & 
    \boldsymbol{\mathcal{J}}_{\tau_{\text{p},xy}} & 
    \boldsymbol{\mathcal{J}}_{\tau_{\text{p},xz}} & 
    \boldsymbol{\mathcal{J}}_{\tau_{\text{p},yz}}
  \end{pmatrix} 
  \cdot
  \begin{pmatrix}
    \boldsymbol{\zeta}_p \\[0.2em] \boldsymbol{\zeta}_u \\[0.2em] \boldsymbol{\zeta}_v \\[0.2em] \boldsymbol{\zeta}_w \\[0.2em] \boldsymbol{\zeta}_{\tau_{\text{p},xx}} \\[0.2em] \boldsymbol{\zeta}_{\tau_{\text{p},yy}} \\[0.2em] \boldsymbol{\zeta}_{\tau_{\text{p},zz}} \\[0.2em] \boldsymbol{\zeta}_{\tau_{\text{p},xy}} \\[0.2em] \boldsymbol{\zeta}_{\tau_{\text{p},xz}} \\[0.2em] \boldsymbol{\zeta}_{\tau_{\text{p},yz}}
  \end{pmatrix}
  = \mathbf{b}, \label{eq:eqsys}
\end{equation}
where $\boldsymbol{\mathcal{A}}_\Gamma$ to $\boldsymbol{\mathcal{J}}_\chi$ are the  $N \times N$ coefficient submatrices of the solution variables $\Gamma$ associated with the continuity equation ($\boldsymbol{\mathcal{A}}$), the three momentum equations ($\boldsymbol{\mathcal{B}}-\boldsymbol{\mathcal{D}}$) and the six constitutive equations of the polymer stress tensor ($\boldsymbol{\mathcal{E}}-\boldsymbol{\mathcal{J}}$). The subvectors $\boldsymbol{\zeta}_\Gamma$ of length $N$ hold the solution of the implicitly sought variables $\Gamma$ and the right-hand side vector $\mathbf{b}$ of length $10N$ holds all known contribution from previous nonlinear iterations and time-steps.
The solution procedure performs nonlinear iterations in which this system of linearized and discretized governing equations, Eq.~\eqref{eq:eqsys}, is solved using the Block-Jacobi pre-conditioner and the BiCGSTAB solver of the software library PETSc \citep{petsc-user-ref,petsc-web-page} until a pre-defined solver tolerance is satisfied. Subsequently, the deferred quantities are updated and Eq.~\eqref{eq:eqsys} is solved again. This procedure continues until the conservation error of the nonlinear set of governing conservation laws satisfies a predefined maximum error \citep{Denner2020}, at which point the solution procedure moves to the next time-step.

The Newton linearization of the advection terms in the momentum equations and the constitutive model yields an implicit contribution of the fluxes $F_f^{(n+1)}$. The flux, thus, introduces an implicit pressure and velocity dependency in all governing equations. Furthermore, the implicit treatment of the polymer stress term and the stress-velocity coupling terms in the momentum equations, alongside the implicit treatment of the upper-convected time derivative of the polymer stress tensor and strain-rate tensor in the constitutive model, provides a strong implicit coupling of the velocity field and the polymer stress tensor. 

\section{Results}
\label{sec:results}

Four representative test cases are considered to demonstrate the capabilities of the proposed numerical framework for single-phase and interfacial flows. First, a lid-driven cavity containing a viscoelastic fluid described by the LPTT model is considered in Section \ref{sec:results_ldc} to assess the basic predictive accuracy of the proposed algorithm. Two-dimensional Taylor vortices are simulated in Section \ref{sec:results_taylor} to quantify the influence of the stress-velocity coupling on the conservation of kinetic energy.
In Section \ref{sec:results_sd}, a Newtonian droplet in a shear-thinning Giesekus fluid is subjected to a shear flow at different Weissenberg numbers, allowing a direct comparison with the results recently reported by \citet{Wang2022b} using a state-of-the-art Lattice-Boltzmann method. A bubble rising in a viscoelastic EPTT fluid under the action of gravity is considered in Section \ref{sec:results_rb}, where we focus particularly on the jump discontinuity in the terminal rise velocity of the bubble and the related negative-wake phenomenon, as studied in detail by \citet{Niethammer2019}.

\subsection{Lid-driven cavity}
\label{sec:results_ldc}
\begin{figure}
  \begin{center}
    \includegraphics[trim=0 0 0 3.5cm, clip=true, width=0.65\linewidth]{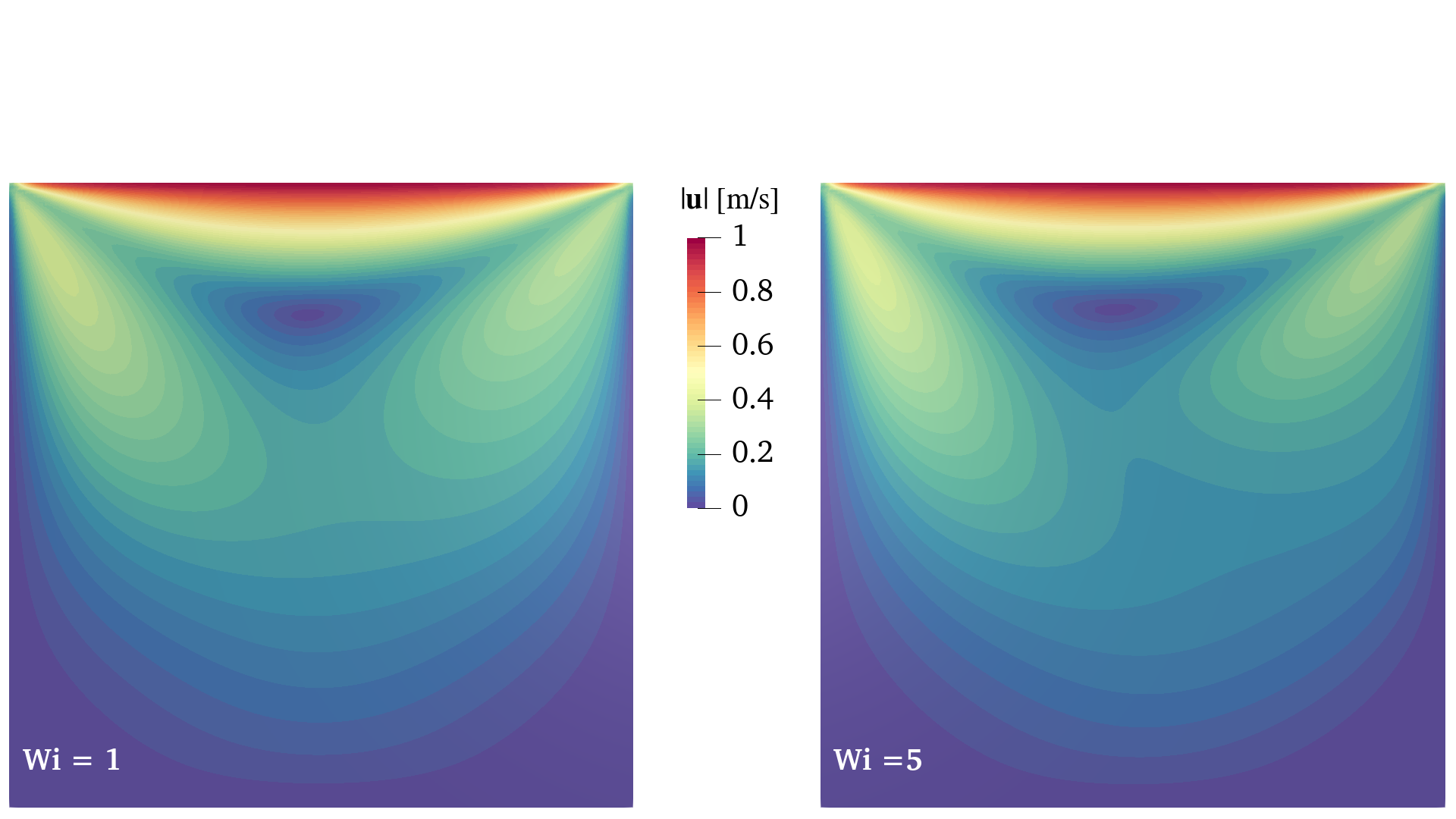}
    \caption{Velocity contours of the considered LPTT fluid in a lid-driven cavity at steady state, for $\mathrm{Wi} \in \{1,5\}$, on an equidistant Cartesian mesh with $160 \times 160$ cells.}
    \label{fig:ldc_umag}
  \end{center}
\end{figure}

A square cavity with edge length $L$ is considered, the top wall of which moves at a constant velocity $U$, with the shear rate defined as $\dot{\gamma} = U/L$. Following \citet{Yapici2012}, we consider an LPTT fluid with $\beta = \mu / (\mu + \eta) = 0.3$, $\epsilon = 0.25$ and $\xi = 0$. Different mesh resolutions ranging from $20 \times 20$ to $160 \times 160$ cells are considered and the applied time-step $\Delta t$ is defined adaptively to correspond to a Courant number of $\mathrm{Co} = \mathbf{u} \, \Delta t/\Delta x \simeq 0.9$. The flow has a Weissenberg number of $\mathrm{Wi}= \dot{\gamma} \lambda \in \{1, 5\}$ and a Reynolds number of $\mathrm{Re} = \rho \dot{\gamma} L^2 / (\mu + \eta) = 10^{-4}$. The contours of the velocity magnitude at steady state of both cases, with $U=1$ m/s, are shown in Figure \ref{fig:ldc_umag}.

\begin{figure}
  \begin{center}
  \subfloat[$u$-velocity]{\includegraphics[scale=1]{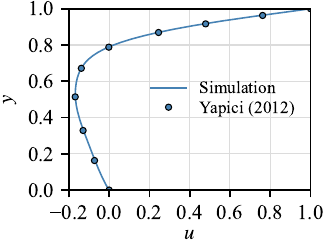}} \hfill
  \subfloat[$v$-velocity]{\includegraphics[scale=1]{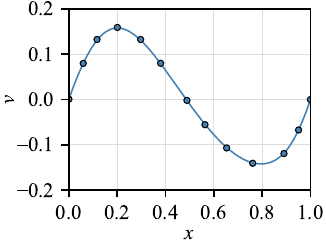}} \hfill
  \subfloat[Mesh convergence]{\includegraphics[scale=1]{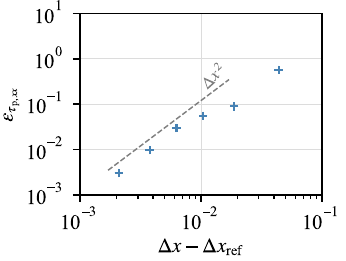}}
  \caption{Velocity profiles along the respective centerlines of the lid-driven cavity, obtained on the reference mesh with $160 \times 160$ cells, and convergence of the error $\varepsilon_{\tau_{\mathrm{p},xx}}$ in normal polymer stress component $\tau_{\mathrm{p},xx}$ relative to the reference mesh, for $\mathrm{Wi} = 1$. The results of \citet{Yapici2012} are shown for reference.}  \label{fig:ldc_Wi1}
  \end{center}
\end{figure}

\begin{figure}
  \begin{center}
  \subfloat[$u$-velocity]{\includegraphics[scale=1]{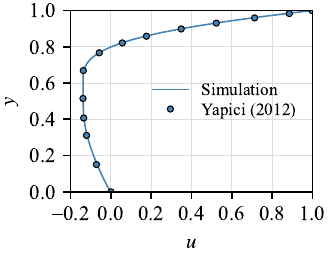}} \hfill
  \subfloat[$v$-velocity]{\includegraphics[scale=1]{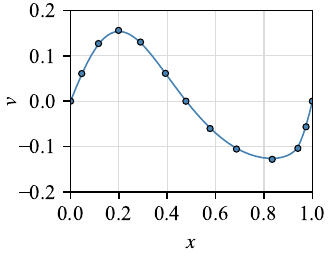}} \hfill
  \subfloat[Mesh convergence]{\includegraphics[scale=1]{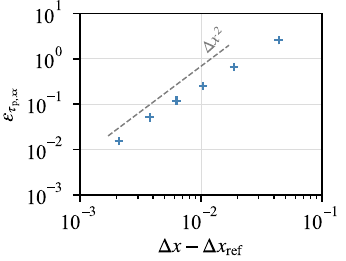}}
  \caption{Velocity profiles along the respective centerlines of the lid-driven cavity, obtained on the reference mesh with $160 \times 160$ cells, and covergence of the error $\varepsilon_{\tau_{\mathrm{p},xx}}$ in normal polymer stress component $\tau_{\mathrm{p},xx}$ relative to the reference mesh, for $\mathrm{Wi} = 5$. The results of \citet{Yapici2012} are shown for reference.}  \label{fig:ldc_Wi5}
  \end{center}
\end{figure}

Figures \ref{fig:ldc_Wi1} and \ref{fig:ldc_Wi5} show the velocity profiles along both centerlines, as well as the error $\varepsilon_{\tau_{\mathrm{p},xx}}$ in normal polymer stress component $\tau_{\mathrm{p},xx}$ at $\mathbf{x} = (0.9L \ 0.9L)$, for $\mathrm{Wi} = 1$ and $\mathrm{Wi} = 5$, respectively. For both considered Weissenberg numbers, the velocity profiles are in excellent agreement with the results reported by \citet{Yapici2012} for the same cases and using the same mesh resolution. The error $\varepsilon_{\tau_{\mathrm{p},xx}}$ in normal polymer stress component $\tau_{\mathrm{p},xx}$ converges, as expected, with second order compared to the solution on the finest mesh, if the mesh resolution is sufficiently high.

\subsection{Taylor vortices}
\label{sec:results_taylor}

\begin{figure}
  \begin{center}
  \includegraphics[width=\linewidth]{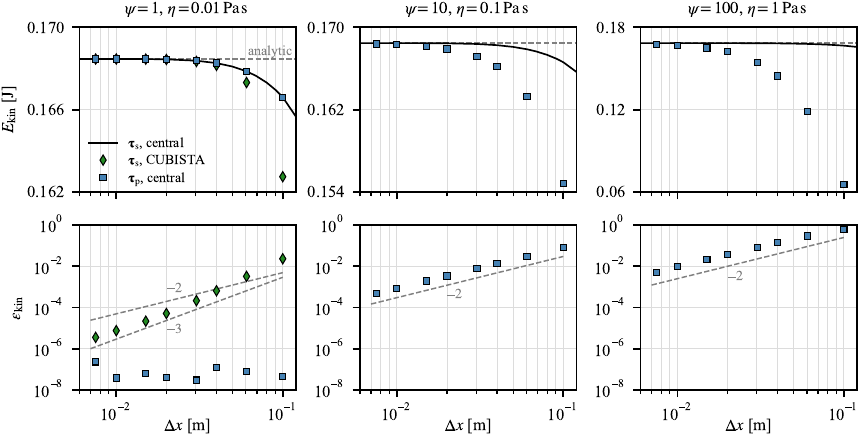}
  \caption{Results of the Newtonian Taylor vortices at $t=1 \mathrm{s}$ for $\mathrm{Re} = 100$, applying $\boldsymbol{\tau}=\boldsymbol{\tau}_\mathrm{s}$, $\boldsymbol{\tau}=\boldsymbol{\tau}_\mathrm{p}$ with the stress-velocity coupling in the momentum equations defined by Eq.~\eqref{eq:momentum_taylor}, for different parameter sets and $\lambda = 0$ For $\boldsymbol{\tau}=\boldsymbol{\tau}_\mathrm{s}$ results obtained with both the central differencing scheme and the CUBISTA scheme are shown, for $\boldsymbol{\tau}=\boldsymbol{\tau}_\mathrm{p}$ only results using the central differencing scheme are shown. Top row: Kinetic energy $E_\mathrm{kin}$ integrated over the domain as a function of mesh spacing $\Delta x$, where the analytic kinetic energy is shown by the dashed line.
  Bottom row: Error $\varepsilon_\mathrm{kin}$, see Eq.~\eqref{eq:tv_kinerror}, as a function of mesh spacing $\Delta x$, incurred when using $\boldsymbol{\tau}=\boldsymbol{\tau}_\mathrm{p}$ compared to $\boldsymbol{\tau}=\boldsymbol{\tau}_\mathrm{s}$, for different parameter sets. }
  \label{fig:tv_newtonian}
  \end{center}
\end{figure}

The evolution of two-dimensional Taylor vortices are simulated to analyze the artificial dissipation of kinetic energy contributed by the stress-velocity coupling. 
With this test case we were able to demonstrate that the fully coupled algorithm for Newtonian flows that is underpinning the proposed algorithm for viscoelastic flows does not introduce numerical diffusion if central differencing is applied \citep{Denner2020}, aside from the numerical diffusion associated with the MWI used for the definition of the fluxes, which, however, decays with $\Delta x^3$.

Following the work of \citet{Ham2004} as well as our previous work \citep{Bartholomew2018, Denner2020}, the computational domain has the dimensions $2\, \mathrm{m} \times 2\, \mathrm{m}$ and is periodic in all directions, such that no boundary conditions need to be considered. For a Newtonian fluid, the velocity and pressure are given as
\begin{align}
  u &= - \cos (\pi x) \, \sin (\pi y) \, \mathrm{e}^{-\frac{2\pi^2 t}{\mathrm{Re}}}\\
  v &= \sin (\pi x) \, \cos (\pi y) \, \mathrm{e}^{-\frac{2\pi^2 t}{\mathrm{Re}}}\\
  p & = - \frac{\cos(2 \pi x) + \cos (2 \pi y)}{4}  \, \mathrm{e}^{-\frac{4\pi^2 t}{\mathrm{Re}}},
\end{align}
from which the initial conditions for the simulations are readily obtained for $t=0$. Integrating the kinetic energy analytically and numerically over the domain $\Omega$ yields for a Newtonian fluid with constant density
\begin{equation}
  E_\mathrm{kin}(t) = \frac{1}{2} \int_\Omega \rho \mathbf{u}(t)^2 \, \mathrm{d}\Omega = \frac{1}{4} \,  \mathrm{e}^{-\frac{4\pi^2 t}{\mathrm{Re}}} \approx \frac{\rho}{2} \sum_f \mathbf{u}_P(t)^2 V_P . \label{eq:kin_taylor}
\end{equation}
The fluid occupying the computational domain has a density of $\rho=1 \, \mathrm{kg/m}^3$ and the time-step applied for all simulations is $\Delta t = 2 \times 10^3 \, \mathrm{s}$. If not stated otherwise, the central differencing scheme is applied for the discretization of the advection terms.

We consider a Newtonian fluid, such that the momentum equations are given as
\begin{equation}
  \rho \left[ \frac{\partial \mathbf{u}}{\partial t} +  \nabla \cdot (\mathbf{u} \otimes \mathbf{u}) \right] = -\nabla p + \nabla \cdot \boldsymbol{\tau}. \label{eq:momentum_taylor}
\end{equation}
However, the stress tensor $\boldsymbol{\tau}$ is now either the solvent stress tensor $\boldsymbol{\tau}_\mathrm{s}$ or the polymer stress tensor $\boldsymbol{\tau}_\mathrm{p}$ under the assumption of $\lambda = 0$, $\alpha=0$, and $\xi=0$, for which the constitutive model reduces to
\begin{equation}
  \tau_\mathrm{p} = \frac{\eta}{\psi} \left(\nabla \mathbf{u} + \nabla \mathbf{u}^\mathrm{T} \right).
\end{equation}
In both scenarios, assuming either $\boldsymbol{\tau}=\boldsymbol{\tau}_\mathrm{s}$ or $\boldsymbol{\tau}=\boldsymbol{\tau}_\mathrm{p}$ in the momentum equations, the results should be identical as long as $\mu = \eta/\psi$. The only difference between the two scenarios is, therefore, the coupling of the polymer stress with the velocity field described in Section \ref{sec:stressvelocitycoupling}, which is not required for the solvent stress. Please note, the constitutive model is solved for the polymer stress to demonstrate the influence of the stress-velocity coupling. As the weighting coefficient for the stress-velocity coupled we use $\hat{\eta}=\eta$, following \citet{Fernandes2019}, as conventionally used in the literature.

Figure \ref{fig:tv_newtonian} shows the kinetic energy integrated over the domain at $t=1 \mathrm{s}$ for $\mathrm{Re} = 100$. Applying $\psi = 1$ for the polymer stress tensor, all cases are in excellent agreement with each other, converging towards the analytical value given by Eq.~\eqref{eq:kin_taylor}. This is to be expected for using the polymer stress in conjunction with the stress-velocity coupling and $\hat{\eta} = \eta$, because in this case the stress-velocity coupling substitutes, in the momentum equations, the large-stencil diffusion term of the strain-rate tensor of the constitutive model by the corresponding small-stencil diffusion term. The errors incurred when using the polymer stress instead of the solvent stress, defined as
\begin{equation}
  \varepsilon_\mathrm{kin}(t) = \frac{|E_{\mathrm{kin,s}}(t) - E_{\mathrm{kin,p}}(t)|}{E_{\mathrm{kin,s}}(t)} ,\label{eq:tv_kinerror}
\end{equation}
where $E_{\mathrm{kin,s}}$ is the kinetic energy obtained using the solvent stress and $E_{\mathrm{kin,p}}$ is the kinetic energy obtained using the polymer stress, are 
numerically negligible. 
Applying the CUBISTA scheme \citep{Alves2003} instead of central differencing to discretise the advection term of the momentum equations also introduces a small amount of numerical diffusion, an error that converges with close to third order under mesh refinement.

Changing the values of the stress function $\psi$ and the polymer viscosity $\eta$ concurrently, such that the ratio $\eta/\psi$ remains unchanged, should yield the same results. However, the stress-velocity coupling imposes a filter on the velocity field that is proportional to $\hat{\eta}$ and $\Delta x^2$, see Eq.~\eqref{eq:velocity_filter}. Figure \ref{fig:tv_newtonian} shows the kinetic energy integrated over the domain at $t=1$ for $\mathrm{Re} = 100$, where $\{\psi = 10, \eta = 0.1 \, \mathrm{Pa \, s}\}$ and $\{\psi = 100, \eta = 1 \, \mathrm{Pa \, s}\}$ for the viscoelastic cases. The polymer stress tensor introduces an error that is, as expected, dependent on the values of $\eta$ and the mesh spacing $\Delta x$. The difference between the results obtained using the polymer stress and using the solvent stress  decays proportional to $\Delta x^2$.

These results demonstrate that the stress-velocity coupling imposes a filter on velocity field and introduces an error that is proportional to $\Delta x^2$, as stipulated by Eq.~\eqref{eq:velocity_filter}. Hence, the employed stress-velocity coupling retains the second-order accuracy of the underlying finite-volume scheme as part of the proposed fully coupled algorithm.

\subsection{Droplet in shear flow}
\label{sec:results_sd}

\begin{figure}
  \begin{center}
\begin{tikzpicture}[scale=0.7]
  \draw [semithick, draw=SkyBlue2!70, fill=SkyBlue2!70] (-3.75,-3.28) rectangle (3.75,3.28);
    \draw[rotate=45, line width = 1.5, color=black, fill=white] (0,0) ellipse (3 and 1.33);
    \draw[line width = 1, -{Stealth[scale=1.1]}] (-0.5, 0) -- (2.5, 0) node[right]{$x$};
    \draw[line width = 1, -{Stealth[scale=1.1]}] (0, -0.5) -- (0, 2.5) node[above]{$z$};
\draw[line width = 1, {Stealth[scale=1.1]}-{Stealth[scale=1.1]}] (-2.8, -3.28) -- (-2.8, 3.28);
\draw (-2.8,0) node[left]{$H$};
\draw[line width = 1.2, -{Stealth[scale=1.1]}] (-1,3.75) -- (1,3.75) node[right]{$U$};
    \draw[line width = 1.2, -{Stealth[scale=1.1]}] (1,-3.75) -- (-1,-3.75) node[left]{$U$};
    \draw[line width = 1, -{Stealth[scale=1.1]}] (0,0) -- (2.12, 2.12);
    \draw[line width = 1, -{Stealth[scale=1.1]}] (0,0) -- (-0.93,0.93);
\draw[line width = 1.5] (-3.75,3.3) -- (3.75, 3.3);
\draw[line width = 1.5] (-3.75,-3.3) -- (3.75, -3.3);
    \draw (1,1.45) node{$\mathbf{L}$};
    \draw (-0.7,0.23) node{$\mathbf{S}$};
    \draw (0.8,0.35) node{$\Theta$};
    \draw [semithick] (1.2,0) arc(0:45:1.2);
\end{tikzpicture}
\caption{Schematic illustration of the droplet in shear flow, where the blue color depicts the viscoelastic fluid.}
\label{fig:sheardrop_schematic}
\end{center}
\end{figure}
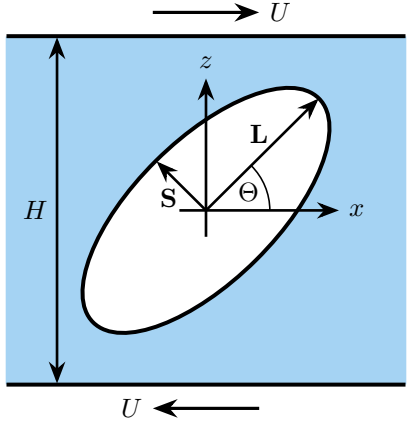

The capabilities of the proposed fully coupled algorithm with respect to interfacial flows with surface tension is demonstrated with a Newtonian droplet situated in a Giesekus fluid between two infinite parallel plates, subject to a shear flow with shear rate $\dot \gamma = 2 U / H$, as illustrated in Figure \ref{fig:sheardrop_schematic}. Following the recent work of \citet{Wang2022b}, the initially spherical droplet with radius $R$ is placed at the center of the computational domain with dimensions $9R \times 5.5 R \times 4R$, represented with an equidistant Cartesian mesh. The plates are modelled as no-slip walls, whereas periodicity is assigned to all other domain boundaries. The host fluid is characterized by a solvent viscosity ratio of $\beta = \mu_\text{h} / (\mu_\text{h} + \eta_\text{h}) = 0.5$, a non-affine parameter of $\xi =0$, and a mobility parameter of $\alpha = 0.3$, the droplet viscosity ratio is $ m = \mu_\text{d} /  (\mu_\text{h} + \eta_\text{h}) = 1$ and the surface tension coefficient $\sigma$ of the fluid interface follows from the considered capillary number $\text{Ca} = {(\mu_\text{h}+\eta_\text{h}) \dot \gamma R}/{\sigma} \in \{ 0.15, 0.25 \}$. The shear flow is in the creeping flow regime, with a Reynolds number of $\text{Re} = {\rho_\text{h} \dot \gamma R^2}/{(\mu_\text{h}+\eta_\text{h})} = 0.1$.

\begin{figure}
  \begin{center}

  \begin{minipage}[b]{0.475\linewidth}
    \centering
    \subfloat[$\text{Ca}=0.15$ and $\text{Wi}=1$]{%
      \includegraphics[width=\linewidth]{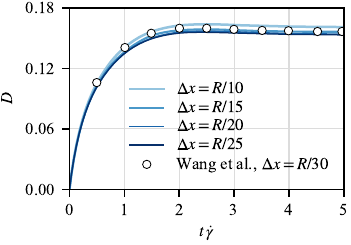}%
      \label{fig:SD_dx}%
    }
  \end{minipage}\hfill
  \begin{minipage}[b]{0.475\linewidth}
    \centering
    \subfloat[$\text{Ca}=0.25$ and mesh resolution of $\Delta x=R/20$]{%
      \includegraphics[width=\linewidth]{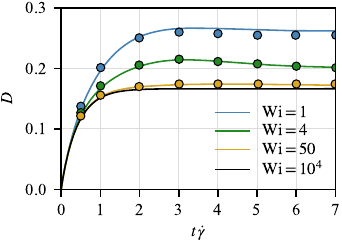}%
      \label{fig:SD_Wi}%
    }
  \end{minipage}

  \caption{Evolution of the Taylor deformation parameter $D$, see Eq.~\eqref{eq:taylor-def}, of the droplet in viscoelastic shear flow. (a) $\text{Ca}=0.15$ and $\text{Wi}=1$, obtained with different mesh resolutions; the circles show the reference results of \citet{Wang2022b}. (b) $\text{Ca}=0.25$ using a mesh resolution of $\Delta x=R/20$, for different Weissenberg numbers $\text{Wi}=\dot \gamma \lambda$; the colored circles show the corresponding reference results of \citet{Wang2022b}.}
  \label{fig:SD}

  \end{center}
\end{figure}

In order to test the mesh convergence of the proposed numerical framework, we consider a droplet with $\text{Ca}= 0.15$ in viscoelastic shear flow with $\text{Wi} = \dot \gamma \lambda = 1$, using different mesh resolutions. 
Figure \ref{fig:SD_dx} shows the evolution of the Taylor deformation parameter,
\begin{equation}
  D = \frac{|\mathbf{L}|-|\mathbf{S}|}{|\mathbf{L}|+|\mathbf{S}|}, \label{eq:taylor-def}
\end{equation}
where $\mathbf{L}$ and $\mathbf{S}$ are the semi-major and semi-minor axes of the deformed droplet, as illustrated in Figure \ref{fig:sheardrop_schematic}.
Even for a mesh resolution as small as 10 cells per initial bubble radius, the proposed numerical framework provides results of reasonable accuracy. The results converge as the mesh resolution is increased and exhibit an overall very good agreement with the reference results of \citet{Wang2022b}.

The evolution of the Taylor deformation parameter $D$, see Eq.~\eqref{eq:taylor-def}, for a droplet with $\text{Ca}=0.25$ and different Weissenberg numbers $\text{Wi}$, using a mesh resolution of $\Delta x = R/20$, is shown in Figure \ref{fig:SD_Wi}. The results obtained with the proposed numerical framework are in excellent agreement with the results of \citet{Wang2022b} up to $\text{Wi}=50$, which is the largest Weissenberg number considered by \citet{Wang2022b}. Even for a strongly elastic case, with $\text{Wi} = 10^4$, the proposed numerical framework is seen to produce physically meaningful results, converging robustly without requiring any form of underrelaxation or log-transformation.

\subsection{Rising bubble}
\label{sec:results_rb}

To further demonstrate the capabilities of the proposed fully coupled algorithm with respect to interfacial flows, a benchmark case of a single gas bubble rising in a viscoelastic liquid is considered. It is a compelling validation case, since experimental studies show a jump in the terminal rise velocity beyond a critical bubble volume and the formation of a negative wake region behind the trailing end of the bubble. The main objective is to achieve quantitative agreement between numerical predictions and experimental measurements and to analyze the flow structures surrounding the bubble to clarify the interplay between rise velocity, bubble deformation, and viscoelastic stress.

\subsubsection{Case setup and overview}

\begin{figure}[htbp]
  \begin{center}
    \includegraphics[trim=0 0 0 0, clip=true, width=0.515\linewidth]{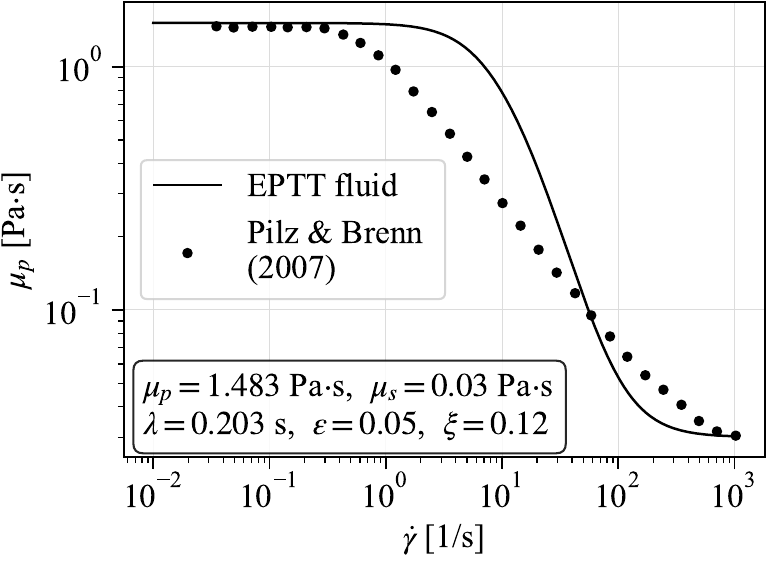}
    \caption{Polymer viscosity as a function of shear rate of the aqueous 0.8\% weight P2500 solution of \citet{Pilz2007} with fitted data for the EPTT model.}
    \label{fig:EPPT fit}
  \end{center}
\end{figure}

The experimental results of a rising bubble in an aqueous P2500 0.8\% weight viscoelastic liquid of  \citet{Pilz2007} are used to validate the proposed fully coupled algorithm. As illustrated in Figure \ref{fig:EPPT fit}, the polymer solution is shear thinning, hence the exponential Phan-Thien Tanner (EPTT) model is used and the model parameters are determined by fitting the viscosity material function of the used model in steady shear flow to experimental rheology data. For the complete derivation of the viscosity material function the reader is referred to \citet{Alves2001}. Liquid density, polymer viscosity, solvent viscosity, surface tension coefficient, and relaxation time are given by \citet{Pilz2007}, while the extensibility coefficient and the slip parameter are determined from the fitting procedure as illustrated in Figure \ref{fig:EPPT fit}. The resultant model parameters are shown in Table \ref{tab:material_parameters} following the work of \citet{Niethammer2019}.

\begin{table}[t]
  \centering
  \caption{Material parameters of P2500 0.8\% weight aqueous viscoelastic liquid.}
  \label{tab:material_parameters}
  \begin{tabular}{%
      S[table-format=4.1]
      S[table-format=1.3]
      S[table-format=1.2]
      S[table-format=1.3]
      S[table-format=1.5]
      S[table-format=1.2]
      S[table-format=1.1]
  }
    \toprule
    {$\rho_l$ [\si{\kilogram\per\meter\cubed}]} &
    {$\eta$ [\si{\pascal\second}]} &
    {$\mu$ [\si{\pascal\second}]} &
    {$\lambda$ [\si{\second}]} &
    {$\sigma$ [\si{\newton\per\meter}]} &
    {$\xi$} &
    {$\epsilon$} \\
    \midrule
    1000.9 & 1.483 & 0.03 & 0.203 & 0.07555 & 0.12 & 0.05 \\
    \bottomrule
  \end{tabular}
\end{table}

The initially spherical bubble with diameter $D$ is placed at the center of a cubic domain of size $20D \times 20D \times 20D$ to eliminate the effects of confinement from the domain boundaries, discretized using an equidistant Cartesian mesh. The bubble is rising in positive $y$-axis due to buoyancy. At the domain boundaries, the velocity field is prescribed using a combination of Dirichlet and Neumann conditions. On the $x$ and $z$ boundaries, impermeable free-slip boundaries are applied: the normal velocity component is set to zero (Dirichlet), while homogeneous Neumann conditions are imposed for the tangential components (zero normal gradients). At the lower $y$ boundary, a no-slip wall is enforced by prescribing all velocity components to zero (Dirichlet). 
At the upper $y$ boundary, an outlet condition is used by imposing homogeneous Neumann conditions for all velocity components (zero normal gradients), while the pressure is fixed to a reference value (Dirichlet) to provide a well-defined pressure level. The pressure satisfies homogeneous Neumann conditions on all remaining boundaries. Finally, homogeneous Neumann conditions are applied to the polymer-stress components on all domain faces, enforcing zero normal gradients at the boundaries. 

Unlike the droplet in shear flow, where a linear interpolation of the stress function, polymer viscosity, relaxation time and non-affine parameter is applied at the interface, we employ here a smooth but sharpened nonlinear interpolation of the same polymer properties for the rising bubble simulations. The rise-velocity jump and the development of a negative wake are governed by the localization and intensity of viscoelastic stresses in a thin layer around the bubble. In particular, detailed analyses relate the regime change to polymer stretching along the bubble contour and to where the stored elastic energy is released relative to the bubble equator \cite{Bothe2022}. The present nonlinear interpolation confines intermediate properties to a narrow interfacial band while remaining continuous, thereby better preserving the stress distribution near the interface and improving agreement with the reference results. The polymer properties for the rising bubble simulations are defined as
\begin{align}
\phi(\mathbf{x})&=
\frac{\big[1-\mathcal{I}(\mathbf{x})\big]^{n}\,\phi_{\mathrm{a}}
      +\omega\,\big[\mathcal{I}(\mathbf{x})\big]^{n}\,\phi_{\mathrm{b}}}
     {\big[1-\mathcal{I}(\mathbf{x})\big]^{n}
      +\omega\,\big[\mathcal{I}(\mathbf{x})\big]^{n}} \label{eq:phi_nonlinear_x}\\
\omega &=\left(\frac{1-I_c}{I_c}\right)^{n},
\label{eq:omega_def}
\end{align}
where $\omega$ is a weighting factor that sets the location of the transition in terms of the indicator field, $I_c\in(0,1)$ denotes the indicator value used to define the effective interface within the numerically smeared interfacial region, and $n>0$ is a sharpness exponent controlling the steepness of the transition (larger $n$ yields a more abrupt switch between phases, while smaller $n$ produces a smoother variation). In the present work, we use $I_c=0.47$ and $n=20$ for all simulated volumes, as these values provided the sharpest transition and the best agreement with our reference results.

\begin{figure}
  \begin{center}

  \begin{minipage}[b]{0.475\linewidth}
    \centering
    \subfloat[Grid convergence study]{%
      \includegraphics[width=\linewidth]{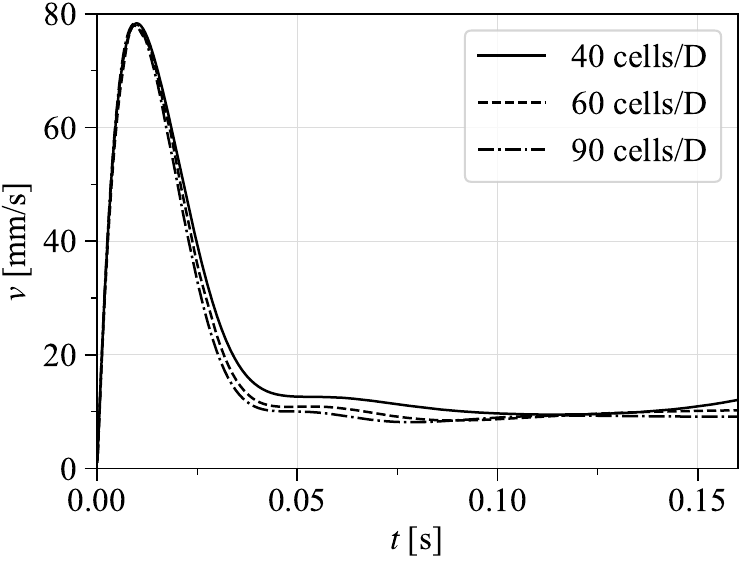}%
      \label{fig:meshconv}%
    }
  \end{minipage}\hfill
  \begin{minipage}[b]{0.475\linewidth}
    \centering
    \subfloat[Time step convergence study]{%
      \includegraphics[width=\linewidth]{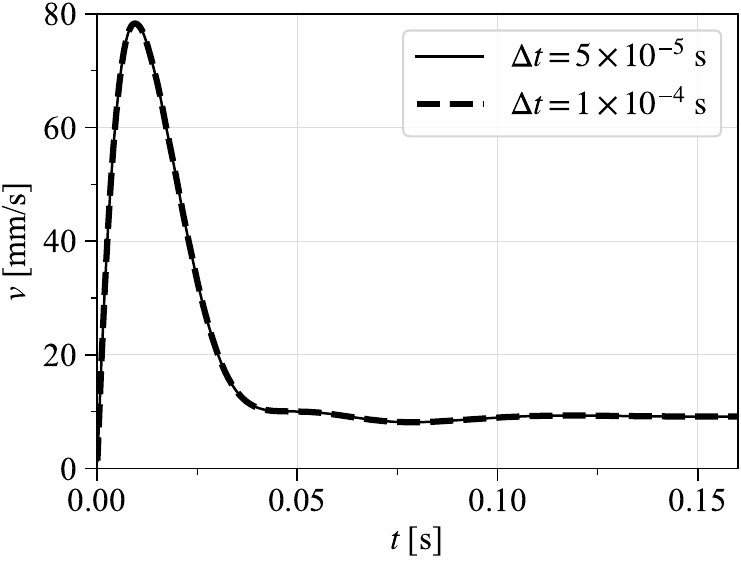}%
      \label{fig:timeconv}%
    }
  \end{minipage}

\caption{Transient rise velocity of a $30\,\mathrm{mm}^3$ bubble as a function of time.
(a) Grid-convergence study using three meshes with different numbers of mesh cells per bubble diameter at $\Delta t=1\times10^{-4}\,\mathrm{s}$. The spatial resolution is increased by a factor of $1.5$ from one mesh level to the next.(b) Time-step convergence study at fixed spatial resolution of 60 mesh cells per bubble diameter.}\label{fig:meshtimeconv}
  \end{center}
\end{figure}

A mesh and time-step convergence study was performed to assess the sensitivity of the numerical results to spatial resolution and temporal discretization. The purpose of this study is to determine an efficient spatial and temporal resolution (mesh cell size and time-step) that still provides reliable predictions of the terminal rise velocity. The bubble rise velocity is defined as the vertical velocity component of the bubble centroid, computed as the indicator weighted volume average of the velocity field. Numerical results for a bubble volume of $30\,\mathrm{mm}^3$ are compared across three mesh resolutions and two time-step sizes to assess grid and time-step convergence, as illustrated in Figures \ref{fig:meshconv} and \ref{fig:timeconv}, respectively. Adaptive mesh refinement is applied in the vicinity of the bubble, providing high resolution around the interface while allowing for a coarser mesh in regions farther from the bubble. Figures~\ref{fig:meshconv} and \ref{fig:timeconv} show that a resolution of $60$ mesh cells per bubble diameter and a time-step of $1\times10^{-4}$ s are sufficient, as further refinement or a smaller time-step yield negligible changes in the results. The time step is selected such that it always satisfies the CFL stability constraint of 0.2.
  
\subsubsection{Results}

Building on the numerical settings and material properties established in the previous section, we report the transient rise velocity evolution and the terminal rise velocity as a function of bubble volume, and discuss the associated negative wake appearance over a region downstream the trailing end of the supercritical bubbles.

\begin{figure}
  \begin{center}

  \begin{minipage}[b]{0.475\linewidth}
    \centering
    \subfloat[Transient rise velocity]{%
      \includegraphics[width=\linewidth]{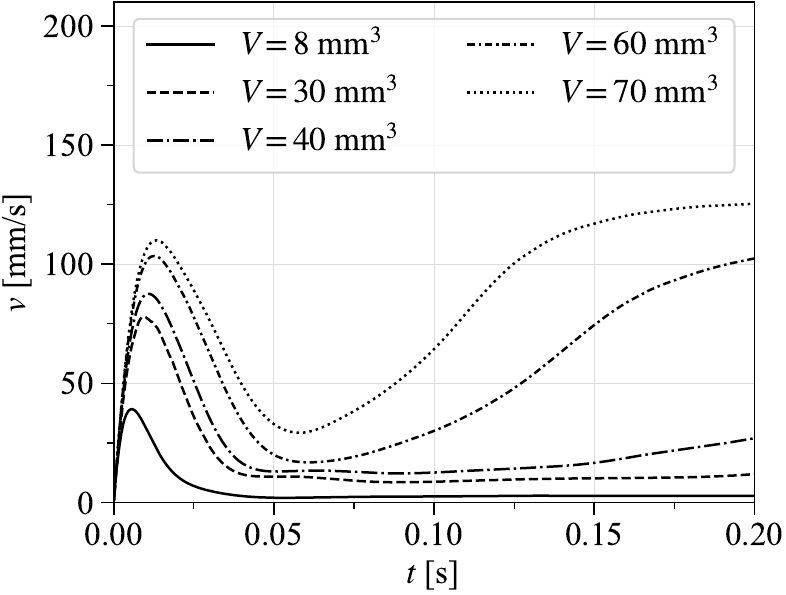}%
      \label{fig:transientvelocity}%
    }
  \end{minipage}\hfill
  \begin{minipage}[b]{0.475\linewidth}
    \centering
    \subfloat[Terminal rise velocity]{%
      \includegraphics[width=\linewidth]{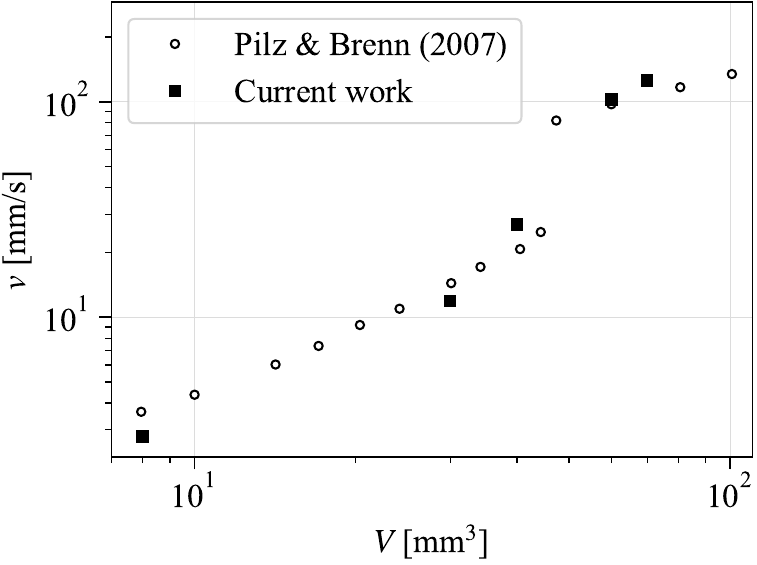}%
      \label{fig:terminalvelocity}%
    }
  \end{minipage}

\caption{Transient and terminal rise velocity of different bubble volumes as a function of time.
(a) Transient rise velocity of five bubble volumes $8$, $30$, $40$, $60$, and $70\,\mathrm{mm}^3$. (b) Terminal rise velocity of five bubble volumes $8$, $30$, $40$, $60$, and $70\,\mathrm{mm}^3$ compared to the experiments of \citet{Pilz2007}.}
  \end{center}
\end{figure}

Figure~\ref{fig:transientvelocity} shows the transient rise velocity of four bubble volumes as a function of time, including three subcritical bubble volumes, $V<V_c$ ($8$, $30$, and $40$ $~\mathrm{mm}^3$) and two supercritical bubble volumes, $V>V_c$ ($60$ and $70~\mathrm{mm}^3$). Based on their experimental measurements, \citet{Pilz2007} reported a critical volume of $V_c \approx 46~\mathrm{mm}^3$ for the considered fluid combination. In the subcritical regime, the bubbles accelerate rapidly from rest and reach a single local maximum at early times, followed by a gradual deceleration toward a steady terminal value. The approach to the terminal state is monotonic after the first maximum and no secondary acceleration stage is observed. In contrast, the supercritical cases exhibit a distinctly different two-stage transient response. The initial acceleration and first local maximum are similar to the subcritical cases, however the subsequent deceleration continues past the terminal value and leads to a local minimum. After this minimum, the bubble undergoes a second acceleration stage and eventually reaches a terminal velocity that is higher than the first local maximum in the considered cases. This non-monotonic evolution is the characteristic signature of the supercritical regime. Figure~\ref{fig:terminalvelocity} reports the terminal rise velocity as a function of bubble volume, comparing the present simulation results with the measurements of \citet{Pilz2007}. The results capture the jump discontinuity in terminal velocity and show good overall agreement with the experimental data.

\begin{figure}
  \centering

  \begin{minipage}[b]{0.475\linewidth}
    \centering
    \includegraphics[width=\linewidth]{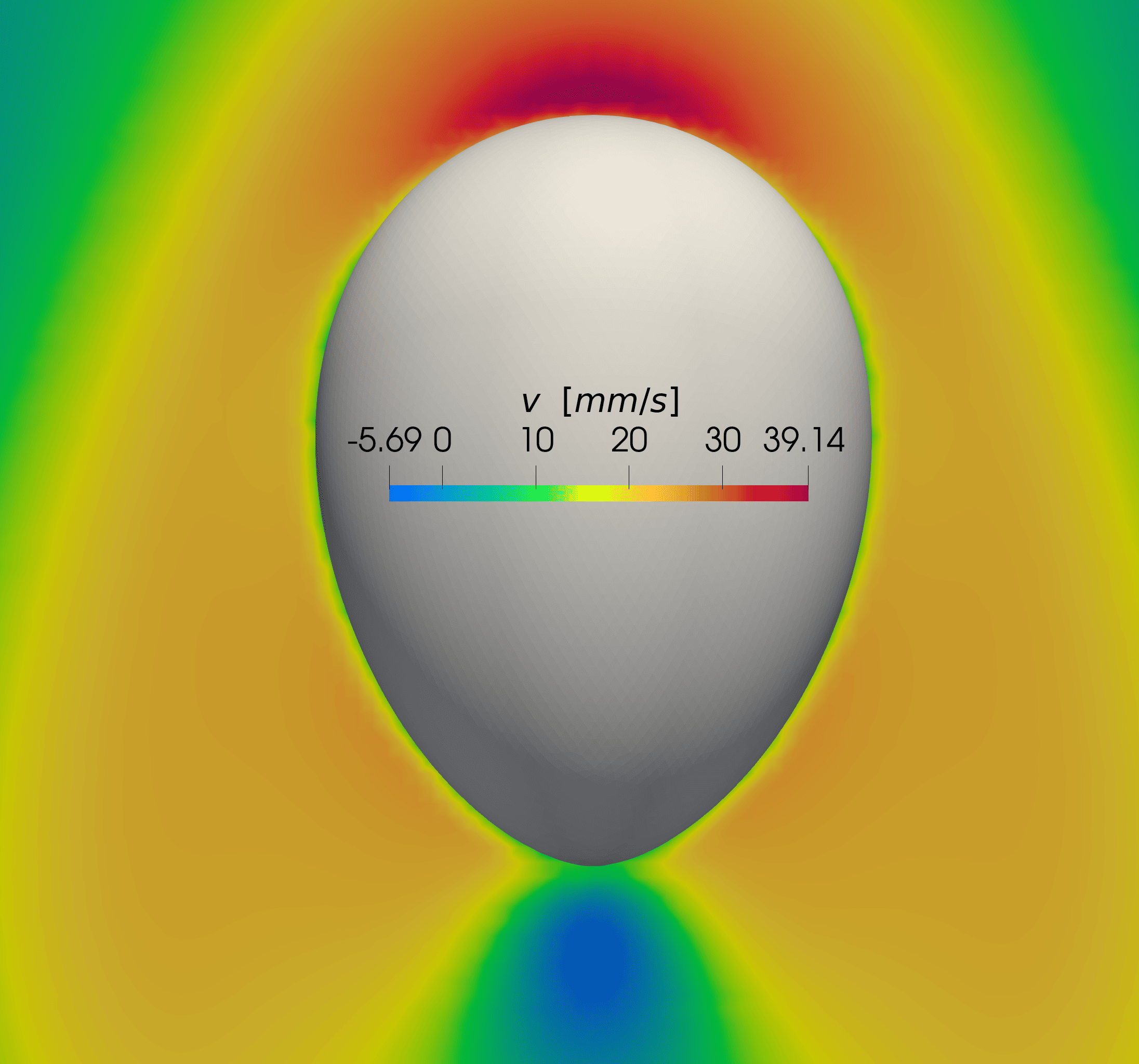}
    \vspace{-0.3em}
    {\small (a) $v$ at $t=0.05\,\mathrm{s}$}
  \end{minipage}\hfill
  \begin{minipage}[b]{0.475\linewidth}
    \centering
    \includegraphics[width=\linewidth]{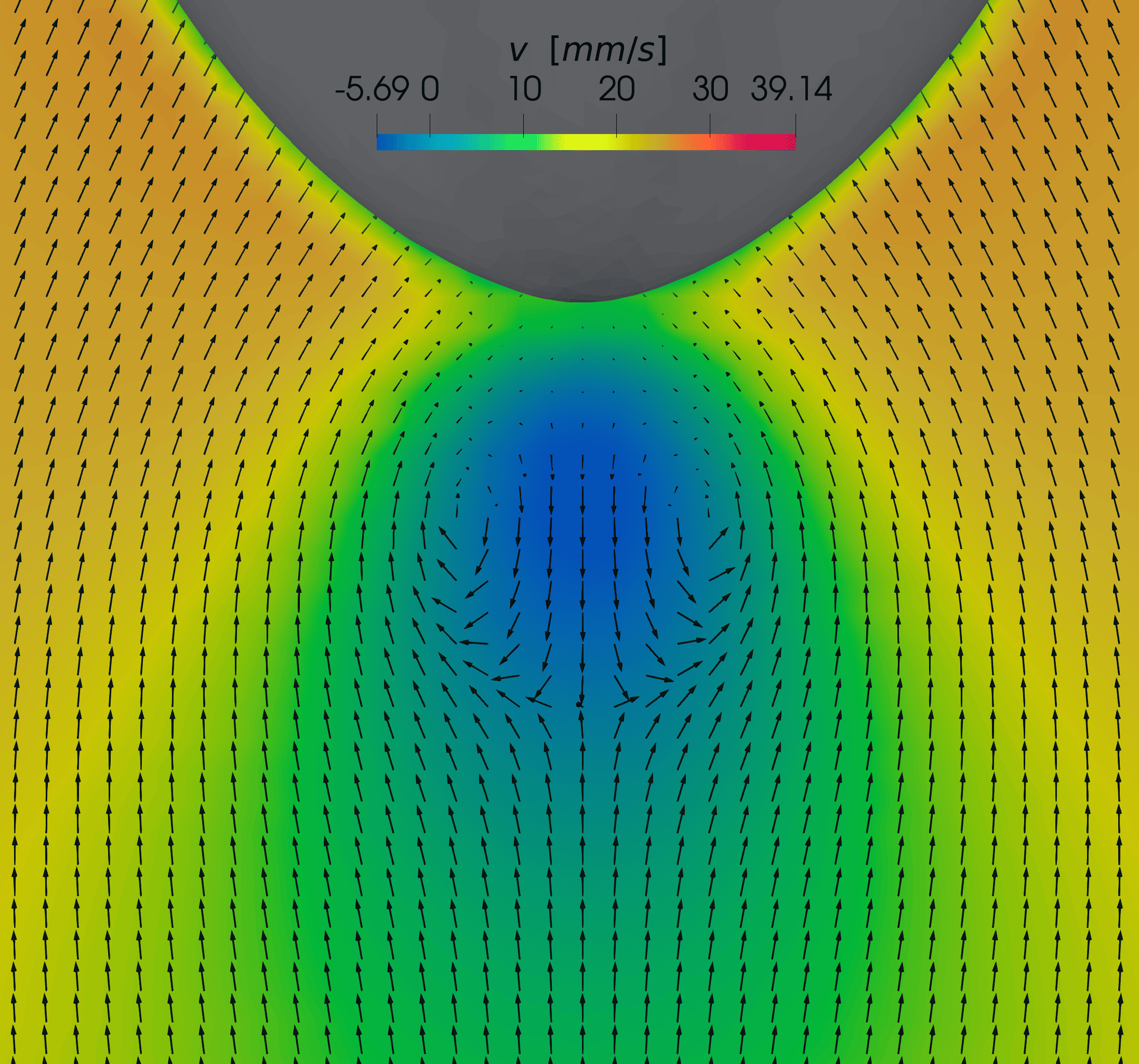}
    \vspace{-0.3em}
    {\small (b) Velocity vectors at $t=0.05\,\mathrm{s}$ (direction only)}
  \end{minipage}

  \vspace{0.8em}

  \begin{minipage}[b]{0.475\linewidth}
    \centering
    \includegraphics[width=\linewidth]{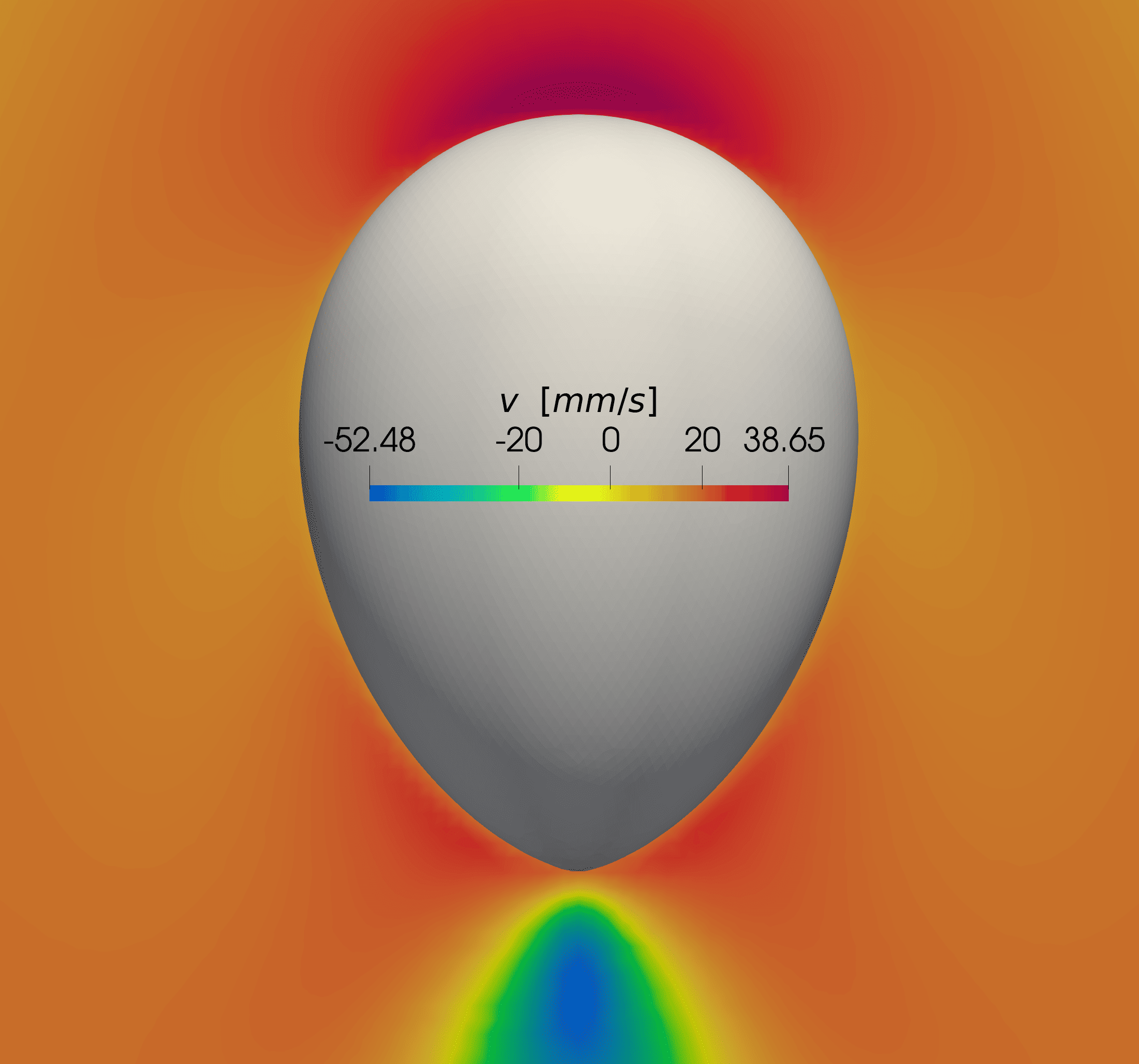}
    \vspace{-0.3em}
    {\small (c) $v$ at $t=0.07\,\mathrm{s}$}
  \end{minipage}\hfill
  \begin{minipage}[b]{0.475\linewidth}
    \centering
    \includegraphics[width=\linewidth]{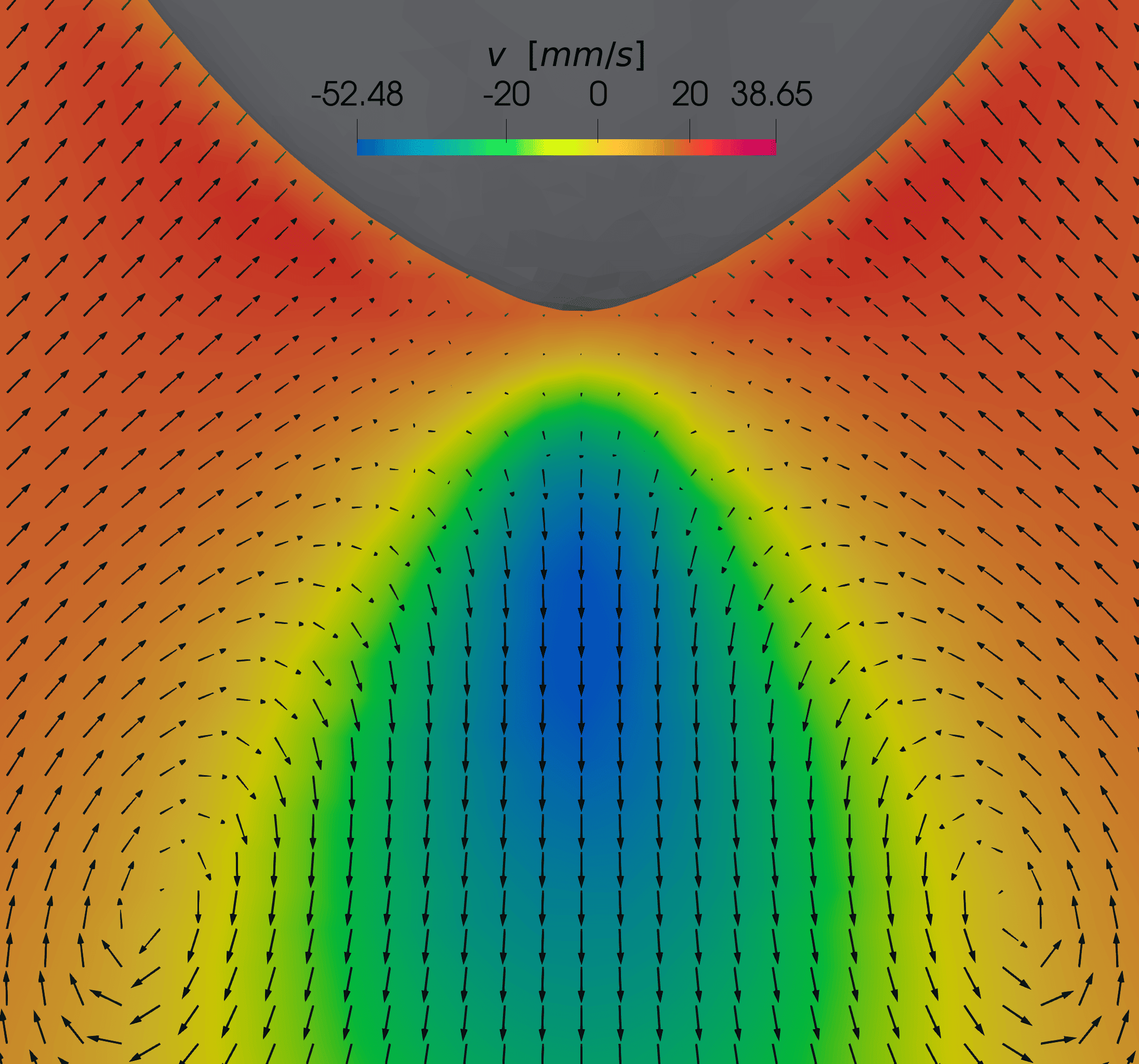}
    \vspace{-0.3em}
    {\small (d) Velocity vectors at $t=0.07\,\mathrm{s}$ (direction only)}
  \end{minipage}

  \caption{Vertical velocity component $v$ (left) and velocity vector field (right) for the bubble with a volume of $70\,\mathrm{mm}^3$ at two representative times. At $t=0.05\,\mathrm{s}$ the onset of reverse flow appears downstream of the trailing end, while at $t=0.07\,\mathrm{s}$ a clearer reversed-flow region is established. Vector glyphs are scaled for visualization of direction only.}
  \label{fig:70mm_wake_panels}
\end{figure}

In the supercritical regime, a negative wake develops behind the bubble. In the inertial frame, the vertical liquid velocity in the bubble's wake reverses direction over a region downstream of the trailing end. Figure~\ref{fig:70mm_wake_panels} shows the wake using vertical velocity contours to identify the onset and spatial growth of the negative wake of the $70\,\mathrm{mm}^3$ bubble at two time instances, $t=0.05\,\mathrm{s}$ where the onset of the negative wake and at $t=0.07\,\mathrm{s}$ where the negative wake is more developed and apparent. In agreement with prior numerical studies, the negative wake develops as a dynamical wake structure that grows downstream of the trailing tip as the bubble approaches its supercritical steady state.

\section{Conclusions}
\label{sec:conclusion}

In this work, we present a fully coupled implicit finite-volume framework for incompressible viscoelastic interfacial flows, in which the continuity equation, momentum equations, and an upper-convected Maxwell constitutive model, including limited extensibility and shear-thinning behaviour, are solved simultaneously for pressure, velocity, and the polymer stress tensor. The proposed discretization treats all relevant couplings implicitly, including the stress-velocity coupling and the pressure-velocity coupling, yielding a tightly coupled linear system at each nonlinear iteration in a standard finite-volume framework. In addition, a state-of-the-art front-tracking approach is used to represent the evolving interface and account for surface tension forces.

The method is assessed using representative single-phase and multiphase benchmarks. For the lid-driven cavity flow with an LPTT fluid, the predicted velocity profiles show excellent agreement with the reference data and stress fields show a second-order mesh convergence when compared against the finest grid. The Taylor vortices test case isolates the numerical impact of the stress-velocity coupling by tracking the domain integrated kinetic energy. The direct substitution parameter set recovers the solvent-stress baseline and converges to the analytical value with central differencing, while scaled stress function and polymer viscosity cases show a residual that decays at second order under mesh refinement, confirming that the coupling preserves second-order accuracy of the current finite-volume discretization within the proposed fully coupled algorithm. For a Newtonian droplet sheared in a shear-thinning Giesekus fluid, the predicted transient Taylor parameter deformation agrees closely with the reference results up to the highest reported Weissenberg number, and remains stable and physically consistent even when extended to substantially higher Weissenberg number. This benchmark therefore supports both interfacial accuracy and robustness with increasing elasticity, without requiring underrelaxation or a log-transformation. Finally, for a bubble rising in an EPTT viscoelastic liquid, the simulations clearly distinguish the transient rise dynamics in the subcritical and supercritical regimes. The predicted terminal rise velocity as a function of bubble volume reproduces the experimentally observed jump discontinuity, while the flow fields show the onset and downstream development of a negative wake in the supercritical case. A grid and time step sensitivity study confirms that these rise velocity predictions are numerically well resolved at the chosen resolution. This benchmark is particularly demanding because it combines strong interface deformation, sharp localization of viscoelastic stresses near the interface, and a regime transition that is highly sensitive to stress-flow coupling. Capturing all of these features consistently highlights the robustness and effectiveness of the fully implicit coupled formulation for challenging viscoelastic interfacial dynamics.

Overall, the four test cases demonstrate that the proposed fully implicit coupled formulation delivers accurate and robust solutions from single phase viscoelastic flows to strongly elastic interfacial dynamics, while maintaining robustness at high Weissenberg numbers without resorting to a log-conformation method.

\section*{Data Availability Statement}
The data that support the findings of this study are reproducible and data is openly available in the
repository with DOI 10.5281/zenodo.18547566, available at \href{https://doi.org/10.5281/zenodo.18547566}{https://doi.org/10.5281/zenodo.18547566}

\section*{Declaration of Generative AI and AI-assisted technologies in the writing process}

The authors used OpenAI's ChatGPT to assist with editing and proofreading this manuscript. The authors then reviewed and revised the text as necessary and assume full responsibility for the final content of the publication.

\section*{Acknowledgements}
We thank Christian Gorges, Fabien Evrard, and Bruno Blais for fruitful discussions. This research was funded by the Deutsche Forschungsgemeinschaft (DFG, German Research Foundation), grant numbers 420239128 and 458610925, and by the Natural Sciences and Engineering Research Council of Canada (NSERC), funding reference number RGPIN-2024-04805.

\bibliographystyle{model1-num-names}

\begin{thebibliography}{86}
\expandafter\ifx\csname natexlab\endcsname\relax\def\natexlab#1{#1}\fi
\providecommand{\bibinfo}[2]{#2}
\ifx\xfnm\relax \def\xfnm[#1]{\unskip,\space#1}\fi
\bibitem[{Williamson et~al.(1997)Williamson, Walters, Bates, Coy, and
  Milton}]{Williamson1997}
\bibinfo{author}{B.~Williamson}, \bibinfo{author}{K.~Walters},
  \bibinfo{author}{T.~Bates}, \bibinfo{author}{R.~Coy},
  \bibinfo{author}{A.~Milton},
\newblock \bibinfo{title}{The viscoelastic properties of multigrade oils and
  their effect on journal-bearing characteristics},
\newblock \bibinfo{journal}{Journal of Non-Newtonian Fluid Mechanics}
  \bibinfo{volume}{73} (\bibinfo{year}{1997}) \bibinfo{pages}{115--126}.
\bibitem[{Gamaniel et~al.(2021)Gamaniel, Dini, and Biancofiore}]{Gamaniel2021}
\bibinfo{author}{S.~Gamaniel}, \bibinfo{author}{D.~Dini},
  \bibinfo{author}{L.~Biancofiore},
\newblock \bibinfo{title}{The effect of fluid viscoelasticity in lubricated
  contacts in the presence of cavitation},
\newblock \bibinfo{journal}{Tribology International} \bibinfo{volume}{160}
  (\bibinfo{year}{2021}) \bibinfo{pages}{107011}.
\bibitem[{Wei et~al.(2019)Wei, Liang, Mei, Yang, and Chen}]{Wei2019b}
\bibinfo{author}{K.~Wei}, \bibinfo{author}{D.~Liang}, \bibinfo{author}{M.~Mei},
  \bibinfo{author}{X.~Yang}, \bibinfo{author}{L.~Chen},
\newblock \bibinfo{title}{A viscoelastic model of compression and relaxation
  behaviors in preforming process for carbon fiber fabrics with binder},
\newblock \bibinfo{journal}{Composites Part B: Engineering}
  \bibinfo{volume}{158} (\bibinfo{year}{2019}) \bibinfo{pages}{1--9}.
\bibitem[{Yuk and Zhao(2018)}]{Yuk2018}
\bibinfo{author}{H.~Yuk}, \bibinfo{author}{X.~Zhao},
\newblock \bibinfo{title}{A {{New 3D Printing Strategy}} by {{Harnessing
  Deformation}}, {{Instability}}, and {{Fracture}} of {{Viscoelastic Inks}}},
\newblock \bibinfo{journal}{Advanced Materials} \bibinfo{volume}{30}
  (\bibinfo{year}{2018}) \bibinfo{pages}{1704028}.
\bibitem[{Nelson et~al.(2020)Nelson, Kundukad, Wong, Khan, and
  Doyle}]{Nelson2020}
\bibinfo{author}{A.~Z. Nelson}, \bibinfo{author}{B.~Kundukad},
  \bibinfo{author}{W.~K. Wong}, \bibinfo{author}{S.~A. Khan},
  \bibinfo{author}{P.~S. Doyle},
\newblock \bibinfo{title}{Embedded droplet printing in yield-stress fluids},
\newblock \bibinfo{journal}{Proceedings of the National Academy of Sciences}
  \bibinfo{volume}{117} (\bibinfo{year}{2020}) \bibinfo{pages}{5671--5679}.
\bibitem[{Frumkin and Bercovici(2021)}]{Frumkin2021}
\bibinfo{author}{V.~Frumkin}, \bibinfo{author}{M.~Bercovici},
\newblock \bibinfo{title}{Fluidic shaping of optical components},
\newblock \bibinfo{journal}{Flow} \bibinfo{volume}{1} (\bibinfo{year}{2021})
  \bibinfo{pages}{E2}.
\bibitem[{Brun(2022)}]{Brun2022}
\bibinfo{author}{P.-T. Brun},
\newblock \bibinfo{title}{Fluid-{{Mediated Fabrication}} of {{Complex
  Assemblies}}},
\newblock \bibinfo{journal}{JACS Au}  (\bibinfo{year}{2022})
  \bibinfo{pages}{jacsau.2c00427}.
\bibitem[{Lai et~al.(2009)Lai, Wang, Wirtz, and Hanes}]{Lai2009}
\bibinfo{author}{S.~K. Lai}, \bibinfo{author}{Y.-Y. Wang},
  \bibinfo{author}{D.~Wirtz}, \bibinfo{author}{J.~Hanes},
\newblock \bibinfo{title}{Micro- and macrorheology of mucus},
\newblock \bibinfo{journal}{Advanced Drug Delivery Reviews}
  \bibinfo{volume}{61} (\bibinfo{year}{2009}) \bibinfo{pages}{86--100}.
\bibitem[{Siginer(2014)}]{Siginer2014}
\bibinfo{author}{D.~A. Siginer}, \bibinfo{title}{Stability of {{Non-Linear
  Constitutive Formulations}} for {{Viscoelastic Fluids}}}, {{SpringerBriefs}}
  in {{Applied Sciences}} and {{Technology}}, \bibinfo{publisher}{Springer
  International Publishing}, \bibinfo{address}{Cham}, \bibinfo{year}{2014}.
\bibitem[{Dupret et~al.(1985)Dupret, Marchal, and Crochet}]{Dupret1985}
\bibinfo{author}{F.~Dupret}, \bibinfo{author}{J.~Marchal},
  \bibinfo{author}{M.~Crochet},
\newblock \bibinfo{title}{On the consequence of discretization errors in the
  numerical calculation of viscoelastic flow},
\newblock \bibinfo{journal}{Journal of Non-Newtonian Fluid Mechanics}
  \bibinfo{volume}{18} (\bibinfo{year}{1985}) \bibinfo{pages}{173--186}.
\bibitem[{Alves et~al.(2021)Alves, Oliveira, and Pinho}]{Alves2021}
\bibinfo{author}{M.~Alves}, \bibinfo{author}{P.~Oliveira},
  \bibinfo{author}{F.~Pinho},
\newblock \bibinfo{title}{Numerical {{Methods}} for {{Viscoelastic Fluid
  Flows}}},
\newblock \bibinfo{journal}{Annual Review of Fluid Mechanics}
  \bibinfo{volume}{53} (\bibinfo{year}{2021}) \bibinfo{pages}{509--541}.
\bibitem[{Keunings(1986)}]{Keunings1986}
\bibinfo{author}{R.~Keunings},
\newblock \bibinfo{title}{On the high {{Weissenberg}} number problem},
\newblock \bibinfo{journal}{Journal of Non-Newtonian Fluid Mechanics}
  \bibinfo{volume}{20} (\bibinfo{year}{1986}) \bibinfo{pages}{209--226}.
\bibitem[{Tsai and Malkus(2000)}]{Tsai2000}
\bibinfo{author}{T.-P. Tsai}, \bibinfo{author}{D.~S. Malkus},
\newblock \bibinfo{title}{Numerical breakdown at high {{Weissenberg}} number in
  non-{{Newtonian}} contraction flows},
\newblock \bibinfo{journal}{Rheologica Acta} \bibinfo{volume}{39}
  (\bibinfo{year}{2000}) \bibinfo{pages}{62--70}.
\bibitem[{Walters and Webster(2003)}]{Walters2003}
\bibinfo{author}{K.~Walters}, \bibinfo{author}{M.~F. Webster},
\newblock \bibinfo{title}{The distinctive {{CFD}} challenges of computational
  rheology},
\newblock \bibinfo{journal}{International Journal for Numerical Methods in
  Fluids} \bibinfo{volume}{43} (\bibinfo{year}{2003})
  \bibinfo{pages}{577--596}.
\bibitem[{Keshtiban et~al.(2004)Keshtiban, Belblidia, and
  Webster}]{Keshtiban2004}
\bibinfo{author}{I.~Keshtiban}, \bibinfo{author}{F.~Belblidia},
  \bibinfo{author}{M.~Webster},
\newblock \bibinfo{title}{Numerical simulation of compressible viscoelastic
  liquids},
\newblock \bibinfo{journal}{Journal of Non-Newtonian Fluid Mechanics}
  \bibinfo{volume}{122} (\bibinfo{year}{2004}) \bibinfo{pages}{131--146}.
\bibitem[{Habla et~al.(2013)Habla, Obermeier, and Hinrichsen}]{Habla2013}
\bibinfo{author}{F.~Habla}, \bibinfo{author}{A.~Obermeier},
  \bibinfo{author}{O.~Hinrichsen},
\newblock \bibinfo{title}{Semi-implicit stress formulation for viscoelastic
  models: {{Application}} to three-dimensional contraction flows},
\newblock \bibinfo{journal}{Journal of Non-Newtonian Fluid Mechanics}
  \bibinfo{volume}{199} (\bibinfo{year}{2013}) \bibinfo{pages}{70--79}.
\bibitem[{Fernandes et~al.(2019)Fernandes, Vuk{\v c}evi{\'c}, Uroi{\'c},
  Simoes, Carneiro, Jasak, and N{\'o}brega}]{Fernandes2019}
\bibinfo{author}{C.~Fernandes}, \bibinfo{author}{V.~Vuk{\v c}evi{\'c}},
  \bibinfo{author}{T.~Uroi{\'c}}, \bibinfo{author}{R.~Simoes},
  \bibinfo{author}{O.~Carneiro}, \bibinfo{author}{H.~Jasak},
  \bibinfo{author}{J.~N{\'o}brega},
\newblock \bibinfo{title}{A coupled finite volume flow solver for the solution
  of incompressible viscoelastic flows},
\newblock \bibinfo{journal}{Journal of Non-Newtonian Fluid Mechanics}
  \bibinfo{volume}{265} (\bibinfo{year}{2019}) \bibinfo{pages}{99--115}.
\bibitem[{Fattal and Kupferman(2004)}]{Fattal2004}
\bibinfo{author}{R.~Fattal}, \bibinfo{author}{R.~Kupferman},
\newblock \bibinfo{title}{Constitutive laws for the matrix-logarithm of the
  conformation tensor},
\newblock \bibinfo{journal}{Journal of Non-Newtonian Fluid Mechanics}
  \bibinfo{volume}{123} (\bibinfo{year}{2004}) \bibinfo{pages}{281--285}.
\bibitem[{Fattal and Kupferman(2005)}]{Fattal2005}
\bibinfo{author}{R.~Fattal}, \bibinfo{author}{R.~Kupferman},
\newblock \bibinfo{title}{Time-dependent simulation of viscoelastic flows at
  high {{Weissenberg}} number using the log-conformation representation},
\newblock \bibinfo{journal}{Journal of Non-Newtonian Fluid Mechanics}
  \bibinfo{volume}{126} (\bibinfo{year}{2005}) \bibinfo{pages}{23--37}.
\bibitem[{Becker et~al.(2023)Becker, Rauthmann, Pauli, and
  Knechtges}]{Becker2023}
\bibinfo{author}{F.~Becker}, \bibinfo{author}{K.~Rauthmann},
  \bibinfo{author}{L.~Pauli}, \bibinfo{author}{P.~Knechtges},
\newblock \bibinfo{title}{An eigenvalue-free implementation of the
  log-conformation formulation},
\newblock \bibinfo{journal}{Journal of Non-Newtonian Fluid Mechanics}
  \bibinfo{volume}{322} (\bibinfo{year}{2023}) \bibinfo{pages}{105133}.
\bibitem[{Doherty et~al.(2024)Doherty, Phillips, and Xie}]{Doherty2024}
\bibinfo{author}{W.~Doherty}, \bibinfo{author}{T.~N. Phillips},
  \bibinfo{author}{Z.~Xie},
\newblock \bibinfo{title}{The log-conformation formulation for single- and
  multi-phase axisymmetric viscoelastic flows},
\newblock \bibinfo{journal}{Journal of Computational Physics}
  \bibinfo{volume}{508} (\bibinfo{year}{2024}) \bibinfo{pages}{113014}.
\bibitem[{Hulsen et~al.(2005)Hulsen, Fattal, and Kupferman}]{Hulsen2005}
\bibinfo{author}{M.~A. Hulsen}, \bibinfo{author}{R.~Fattal},
  \bibinfo{author}{R.~Kupferman},
\newblock \bibinfo{title}{Flow of viscoelastic fluids past a cylinder at high
  {{Weissenberg}} number: {{Stabilized}} simulations using matrix logarithms},
\newblock \bibinfo{journal}{Journal of Non-Newtonian Fluid Mechanics}
  \bibinfo{volume}{127} (\bibinfo{year}{2005}) \bibinfo{pages}{27--39}.
\bibitem[{Afonso et~al.(2009)Afonso, Oliveira, Pinho, and Alves}]{Afonso2009}
\bibinfo{author}{A.~Afonso}, \bibinfo{author}{P.~Oliveira},
  \bibinfo{author}{F.~Pinho}, \bibinfo{author}{M.~Alves},
\newblock \bibinfo{title}{The log-conformation tensor approach in the
  finite-volume method framework},
\newblock \bibinfo{journal}{Journal of Non-Newtonian Fluid Mechanics}
  \bibinfo{volume}{157} (\bibinfo{year}{2009}) \bibinfo{pages}{55--65}.
\bibitem[{Martins et~al.(2015)Martins, Oishi, Afonso, and Alves}]{Martins2015}
\bibinfo{author}{F.~Martins}, \bibinfo{author}{C.~Oishi},
  \bibinfo{author}{A.~Afonso}, \bibinfo{author}{M.~Alves},
\newblock \bibinfo{title}{A numerical study of the {{Kernel-conformation}}
  transformation for transient viscoelastic fluid flows},
\newblock \bibinfo{journal}{Journal of Computational Physics}
  \bibinfo{volume}{302} (\bibinfo{year}{2015}) \bibinfo{pages}{653--673}.
\bibitem[{Niethammer et~al.(2018)Niethammer, Marschall, Kunkelmann, and
  Bothe}]{Niethammer2018}
\bibinfo{author}{M.~Niethammer}, \bibinfo{author}{H.~Marschall},
  \bibinfo{author}{C.~Kunkelmann}, \bibinfo{author}{D.~Bothe},
\newblock \bibinfo{title}{A numerical stabilization framework for viscoelastic
  fluid flow using the finite volume method on general unstructured meshes},
\newblock \bibinfo{journal}{International Journal for Numerical Methods in
  Fluids} \bibinfo{volume}{86} (\bibinfo{year}{2018})
  \bibinfo{pages}{131--166}.
\bibitem[{Fernandes(2022)}]{Fernandes2022}
\bibinfo{author}{C.~Fernandes},
\newblock \bibinfo{title}{A {{Fully Implicit Log-Conformation Tensor Coupled
  Algorithm}} for the {{Solution}} of {{Incompressible Non-Isothermal
  Viscoelastic Flows}}},
\newblock \bibinfo{journal}{Polymers} \bibinfo{volume}{14}
  (\bibinfo{year}{2022}) \bibinfo{pages}{4099}.
\bibitem[{Tom{\'e} et~al.(2012)Tom{\'e}, Castelo, Afonso, Alves, and
  Pinho}]{Tome2012}
\bibinfo{author}{M.~Tom{\'e}}, \bibinfo{author}{A.~Castelo},
  \bibinfo{author}{A.~Afonso}, \bibinfo{author}{M.~Alves},
  \bibinfo{author}{F.~Pinho},
\newblock \bibinfo{title}{Application of the log-conformation tensor to
  three-dimensional time-dependent free surface flows},
\newblock \bibinfo{journal}{Journal of Non-Newtonian Fluid Mechanics}
  \bibinfo{volume}{175--176} (\bibinfo{year}{2012}) \bibinfo{pages}{44--54}.
\bibitem[{Niethammer et~al.(2019)Niethammer, Brenn, Marschall, and
  Bothe}]{Niethammer2019}
\bibinfo{author}{M.~Niethammer}, \bibinfo{author}{G.~Brenn},
  \bibinfo{author}{H.~Marschall}, \bibinfo{author}{D.~Bothe},
\newblock \bibinfo{title}{An extended volume of fluid method and its
  application to single bubbles rising in a viscoelastic liquid},
\newblock \bibinfo{journal}{Journal of Computational Physics}
  \bibinfo{volume}{387} (\bibinfo{year}{2019}) \bibinfo{pages}{326--355}.
\bibitem[{Fernandes et~al.(2019)Fernandes, Faroughi, Carneiro, N{\'o}brega, and
  McKinley}]{Fernandes2019a}
\bibinfo{author}{C.~Fernandes}, \bibinfo{author}{S.~Faroughi},
  \bibinfo{author}{O.~Carneiro}, \bibinfo{author}{J.~M. N{\'o}brega},
  \bibinfo{author}{G.~McKinley},
\newblock \bibinfo{title}{Fully-resolved simulations of particle-laden
  viscoelastic fluids using an immersed boundary method},
\newblock \bibinfo{journal}{Journal of Non-Newtonian Fluid Mechanics}
  \bibinfo{volume}{266} (\bibinfo{year}{2019}) \bibinfo{pages}{80--94}.
\bibitem[{Naseer et~al.(2023)Naseer, Ahmed, Izbassarov, and
  Muradoglu}]{Naseer2023}
\bibinfo{author}{H.~U. Naseer}, \bibinfo{author}{Z.~Ahmed},
  \bibinfo{author}{D.~Izbassarov}, \bibinfo{author}{M.~Muradoglu},
\newblock \bibinfo{title}{Dynamics and interactions of parallel bubbles rising
  in a viscoelastic fluid under buoyancy},
\newblock \bibinfo{journal}{Journal of Non-Newtonian Fluid Mechanics}
  (\bibinfo{year}{2023}) \bibinfo{pages}{105000}.
\bibitem[{Xiao et~al.(2017)Xiao, Denner, and {van Wachem}}]{Xiao2017}
\bibinfo{author}{C.-N. Xiao}, \bibinfo{author}{F.~Denner},
  \bibinfo{author}{B.~{van Wachem}},
\newblock \bibinfo{title}{Fully-coupled pressure-based finite-volume framework
  for the simulation of fluid flows at all speeds in complex geometries},
\newblock \bibinfo{journal}{Journal of Computational Physics}
  \bibinfo{volume}{346} (\bibinfo{year}{2017}) \bibinfo{pages}{91--130}.
\bibitem[{Denner et~al.(2020)Denner, Evrard, and {van Wachem}}]{Denner2020}
\bibinfo{author}{F.~Denner}, \bibinfo{author}{F.~Evrard},
  \bibinfo{author}{B.~{van Wachem}},
\newblock \bibinfo{title}{Conservative finite-volume framework and
  pressure-based algorithm for flows of incompressible, ideal-gas and real-gas
  fluids at all speeds},
\newblock \bibinfo{journal}{Journal of Computational Physics}
  \bibinfo{volume}{409} (\bibinfo{year}{2020}) \bibinfo{pages}{109348}.
\bibitem[{Darwish et~al.(2009)Darwish, Sraj, and Moukalled}]{Darwish2009a}
\bibinfo{author}{M.~Darwish}, \bibinfo{author}{I.~Sraj},
  \bibinfo{author}{F.~Moukalled},
\newblock \bibinfo{title}{A coupled finite volume solver for the solution of
  incompressible flows on unstructured grids},
\newblock \bibinfo{journal}{Journal of Computational Physics}
  \bibinfo{volume}{228} (\bibinfo{year}{2009}) \bibinfo{pages}{180--201}.
\bibitem[{Darwish and Moukalled(2014)}]{Darwish2014}
\bibinfo{author}{M.~Darwish}, \bibinfo{author}{F.~Moukalled},
\newblock \bibinfo{title}{A fully coupled {{Navier-Stokes}} solver for fluid
  flow at all speeds},
\newblock \bibinfo{journal}{Numerical Heat Transfer, Part B: Fundamentals}
  \bibinfo{volume}{65} (\bibinfo{year}{2014}) \bibinfo{pages}{410--444}.
\bibitem[{Chen and Przekwas(2010)}]{Chen2010}
\bibinfo{author}{Z.~Chen}, \bibinfo{author}{A.~J. Przekwas},
\newblock \bibinfo{title}{A coupled pressure-based computational method for
  incompressible/compressible flows},
\newblock \bibinfo{journal}{Journal of Computational Physics}
  \bibinfo{volume}{229} (\bibinfo{year}{2010}) \bibinfo{pages}{9150--9165}.
\bibitem[{Denner and {van Wachem}(2014)}]{Denner2014a}
\bibinfo{author}{F.~Denner}, \bibinfo{author}{B.~{van Wachem}},
\newblock \bibinfo{title}{Fully-coupled balanced-force {{VOF}} framework for
  arbitrary meshes with least-squares curvature evaluation from volume
  fractions},
\newblock \bibinfo{journal}{Numerical Heat Transfer Part B: Fundamentals}
  \bibinfo{volume}{65} (\bibinfo{year}{2014}) \bibinfo{pages}{218--255}.
\bibitem[{Denner et~al.(2018)Denner, Xiao, and {van Wachem}}]{Denner2018b}
\bibinfo{author}{F.~Denner}, \bibinfo{author}{C.-N. Xiao},
  \bibinfo{author}{B.~{van Wachem}},
\newblock \bibinfo{title}{Pressure-based algorithm for compressible interfacial
  flows with acoustically-conservative interface discretisation},
\newblock \bibinfo{journal}{Journal of Computational Physics}
  \bibinfo{volume}{367} (\bibinfo{year}{2018}) \bibinfo{pages}{192--234}.
\bibitem[{Denner and {van Wachem}(2022)}]{Denner2022c}
\bibinfo{author}{F.~Denner}, \bibinfo{author}{B.~{van Wachem}},
\newblock \bibinfo{title}{A {{Unified Algorithm}} for {{Interfacial Flows}}
  with {{Incompressible}} and {{Compressible Fluids}}},
\newblock in: \bibinfo{editor}{D.~Zeidan}, \bibinfo{editor}{L.~T. Zhang},
  \bibinfo{editor}{E.~G. Da~Silva}, \bibinfo{editor}{J.~Merker} (Eds.),
  \bibinfo{booktitle}{Advances in {{Fluid Mechanics}}: {{Modelling}} and
  {{Simulations}}}, \bibinfo{publisher}{Springer Nature Singapore},
  \bibinfo{address}{Singapore}, \bibinfo{year}{2022}, pp.
  \bibinfo{pages}{179--208}.
\bibitem[{Gorges et~al.(2022)Gorges, Evrard, {van Wachem}, and
  Denner}]{Gorges2022}
\bibinfo{author}{C.~Gorges}, \bibinfo{author}{F.~Evrard},
  \bibinfo{author}{B.~{van Wachem}}, \bibinfo{author}{F.~Denner},
\newblock \bibinfo{title}{Reducing volume and shape errors in front tracking by
  divergence-preserving velocity interpolation and parabolic fit vertex
  positioning},
\newblock \bibinfo{journal}{Journal of Computational Physics}
  \bibinfo{volume}{457} (\bibinfo{year}{2022}) \bibinfo{pages}{111072}.
\bibitem[{Darwish et~al.(2015)Darwish, Aziz, and Moukalled}]{Darwish2015}
\bibinfo{author}{M.~Darwish}, \bibinfo{author}{A.~Aziz},
  \bibinfo{author}{F.~Moukalled},
\newblock \bibinfo{title}{A {{Coupled Pressure-Based Finite-Volume Solver}} for
  {{Incompressible Two-Phase Flow}}},
\newblock \bibinfo{journal}{Numerical Heat Transfer, Part B: Fundamentals}
  \bibinfo{volume}{67} (\bibinfo{year}{2015}) \bibinfo{pages}{47--74}.
\bibitem[{Pimenta and Alves(2019)}]{Pimenta2019}
\bibinfo{author}{F.~Pimenta}, \bibinfo{author}{M.~Alves},
\newblock \bibinfo{title}{A coupled finite-volume solver for numerical
  simulation of electrically-driven flows},
\newblock \bibinfo{journal}{Computers \& Fluids} \bibinfo{volume}{193}
  (\bibinfo{year}{2019}) \bibinfo{pages}{104279}.
\bibitem[{Hirt and Nichols(1981)}]{Hirt1981}
\bibinfo{author}{C.~Hirt}, \bibinfo{author}{B.~Nichols},
\newblock \bibinfo{title}{Volume of fluid ({{VOF}}) method for the dynamics of
  free boundaries},
\newblock \bibinfo{journal}{Journal of Computational Physics}
  \bibinfo{volume}{39} (\bibinfo{year}{1981}) \bibinfo{pages}{201--225}.
\bibitem[{Osher and Sethian(1988)}]{Osher1988}
\bibinfo{author}{S.~Osher}, \bibinfo{author}{J.~A. Sethian},
\newblock \bibinfo{title}{Fronts {{Propagating}} with {{Curvature-Dependent
  Speed}}: {{Algorithms}} based on the {{Hamilton-Jacobi Formulation}}},
\newblock \bibinfo{journal}{Journal of Computational Physics}
  \bibinfo{volume}{79} (\bibinfo{year}{1988}) \bibinfo{pages}{12--49}.
\bibitem[{Sussman et~al.(1994)Sussman, Smereka, and Osher}]{Sussman1994}
\bibinfo{author}{M.~Sussman}, \bibinfo{author}{P.~Smereka},
  \bibinfo{author}{S.~Osher},
\newblock \bibinfo{title}{A {{Level Set Approach}} for {{Computing Solutions}}
  to {{Incompressible Two-Phase Flow}}},
\newblock \bibinfo{journal}{Journal of Computational Physics}
  \bibinfo{volume}{114} (\bibinfo{year}{1994}) \bibinfo{pages}{146--159}.
\bibitem[{Unverdi and Tryggvason(1992)}]{Unverdi1992}
\bibinfo{author}{S.~Unverdi}, \bibinfo{author}{G.~Tryggvason},
\newblock \bibinfo{title}{A {{Front-Tracking Method}} for {{Viscous}},
  {{Incompressible}}, {{Multi-fluid Flows}}},
\newblock \bibinfo{journal}{Journal of Computational Physics}
  \bibinfo{volume}{100} (\bibinfo{year}{1992}) \bibinfo{pages}{25--37}.
\bibitem[{Tryggvason et~al.(2001)Tryggvason, Bunner, Esmaeeli, Juric,
  {Al-Rawahi}, Tauber, Han, Nas, and Jan}]{Tryggvason2001}
\bibinfo{author}{G.~Tryggvason}, \bibinfo{author}{B.~Bunner},
  \bibinfo{author}{A.~Esmaeeli}, \bibinfo{author}{D.~Juric},
  \bibinfo{author}{N.~{Al-Rawahi}}, \bibinfo{author}{W.~Tauber},
  \bibinfo{author}{J.~Han}, \bibinfo{author}{S.~Nas}, \bibinfo{author}{Y.~Jan},
\newblock \bibinfo{title}{A front-tracking method for the computations of
  multiphase flow},
\newblock \bibinfo{journal}{Journal of Computational Physics}
  \bibinfo{volume}{169} (\bibinfo{year}{2001}) \bibinfo{pages}{708--759}.
\bibitem[{Habla et~al.(2011)Habla, Marschall, Hinrichsen, Dietsche, Jasak, and
  Favero}]{Habla2011}
\bibinfo{author}{F.~Habla}, \bibinfo{author}{H.~Marschall},
  \bibinfo{author}{O.~Hinrichsen}, \bibinfo{author}{L.~Dietsche},
  \bibinfo{author}{H.~Jasak}, \bibinfo{author}{J.~L. Favero},
\newblock \bibinfo{title}{Numerical simulation of viscoelastic two-phase flows
  using {{openFOAM}}{\textregistered}},
\newblock \bibinfo{journal}{Chemical Engineering Science} \bibinfo{volume}{66}
  (\bibinfo{year}{2011}) \bibinfo{pages}{5487--5496}.
\bibitem[{Giliberto and Desjardins(2026)}]{Giliberto2026}
\bibinfo{author}{J.~V. Giliberto}, \bibinfo{author}{O.~Desjardins},
\newblock \bibinfo{title}{A sharp computational method for simulating
  multiphase viscoelastic flows},
\newblock \bibinfo{journal}{Journal of Non-Newtonian Fluid Mechanics}
  \bibinfo{volume}{348} (\bibinfo{year}{2026}) \bibinfo{pages}{105559}.
\bibitem[{Pillapakkam and Singh(2001)}]{Pillapakkam2001}
\bibinfo{author}{{\relax SB}.~Pillapakkam}, \bibinfo{author}{P.~Singh},
\newblock \bibinfo{title}{A level-set method for computing solutions to
  viscoelastic two-phase flow},
\newblock \bibinfo{journal}{Journal of Computational Physics}
  \bibinfo{volume}{174} (\bibinfo{year}{2001}) \bibinfo{pages}{552--578}.
\bibitem[{Kabanemi and Marcotte(2020)}]{Kabanemi2020}
\bibinfo{author}{K.~K. Kabanemi}, \bibinfo{author}{J.-P. Marcotte},
\newblock \bibinfo{title}{A level set method for simulating wrinkling of
  extruded viscoelastic sheets},
\newblock \bibinfo{journal}{Polymer Engineering \& Science}
  \bibinfo{volume}{60} (\bibinfo{year}{2020}) \bibinfo{pages}{1662--1675}.
\bibitem[{Amani et~al.(2020)Amani, Balc{\'a}zar, Naseri, and
  Rigola}]{Amani2020}
\bibinfo{author}{A.~Amani}, \bibinfo{author}{N.~Balc{\'a}zar},
  \bibinfo{author}{A.~Naseri}, \bibinfo{author}{J.~Rigola},
\newblock \bibinfo{title}{A numerical approach for non-{{Newtonian}} two-phase
  flows using a conservative level-set method},
\newblock \bibinfo{journal}{Chemical Engineering Journal} \bibinfo{volume}{385}
  (\bibinfo{year}{2020}) \bibinfo{pages}{123896}.
\bibitem[{Izbassarov and Muradoglu(2015)}]{Izbassarov2015}
\bibinfo{author}{D.~Izbassarov}, \bibinfo{author}{M.~Muradoglu},
\newblock \bibinfo{title}{A front-tracking method for computational modeling of
  viscoelastic two-phase flow systems},
\newblock \bibinfo{journal}{Journal of Non-Newtonian Fluid Mechanics}
  \bibinfo{volume}{223} (\bibinfo{year}{2015}) \bibinfo{pages}{122--140}.
\bibitem[{Wang et~al.(2022)Wang, Wang, and Liu}]{Wang2022b}
\bibinfo{author}{D.~Wang}, \bibinfo{author}{N.~Wang}, \bibinfo{author}{H.~Liu},
\newblock \bibinfo{title}{Droplet deformation and breakup in shear-thinning
  viscoelastic fluid under simple shear flow},
\newblock \bibinfo{journal}{Journal of Rheology} \bibinfo{volume}{66}
  (\bibinfo{year}{2022}) \bibinfo{pages}{585--603}.
\bibitem[{Yue et~al.(2004)Yue, Feng, Liu, and Shen}]{Yue2004}
\bibinfo{author}{P.~Yue}, \bibinfo{author}{J.~J. Feng},
  \bibinfo{author}{C.~Liu}, \bibinfo{author}{J.~Shen},
\newblock \bibinfo{title}{A diffuse-interface method for simulating two-phase
  flows of complex fluids},
\newblock \bibinfo{journal}{Journal of Fluid Mechanics} \bibinfo{volume}{515}
  (\bibinfo{year}{2004}) \bibinfo{pages}{293--317}.
\bibitem[{Yue et~al.(2006)Yue, Zhou, Feng, {Ollivier-Gooch}, and Hu}]{Yue2006}
\bibinfo{author}{P.~Yue}, \bibinfo{author}{C.~Zhou}, \bibinfo{author}{J.~J.
  Feng}, \bibinfo{author}{C.~F. {Ollivier-Gooch}}, \bibinfo{author}{H.~H. Hu},
\newblock \bibinfo{title}{Phase-field simulations of interfacial dynamics in
  viscoelastic fluids using finite elements with adaptive meshing},
\newblock \bibinfo{journal}{Journal of Computational Physics}
  \bibinfo{volume}{219} (\bibinfo{year}{2006}) \bibinfo{pages}{47--67}.
\bibitem[{Rodriguez and Johnsen(2019)}]{Rodriguez2019}
\bibinfo{author}{M.~Rodriguez}, \bibinfo{author}{E.~Johnsen},
\newblock \bibinfo{title}{A high-order accurate five-equations compressible
  multiphase approach for viscoelastic fluids and solids with relaxation and
  elasticity},
\newblock \bibinfo{journal}{Journal of Computational Physics}
  \bibinfo{volume}{379} (\bibinfo{year}{2019}) \bibinfo{pages}{70--90}.
\bibitem[{Zografos et~al.(2020)Zografos, Afonso, Poole, and
  Oliveira}]{Zografos2020}
\bibinfo{author}{K.~Zografos}, \bibinfo{author}{A.~M. Afonso},
  \bibinfo{author}{R.~J. Poole}, \bibinfo{author}{M.~S.~N. Oliveira},
\newblock \bibinfo{title}{A viscoelastic two-phase solver using a phase-field
  approach},
\newblock \bibinfo{journal}{Journal of Non-Newtonian Fluid Mechanics}
  \bibinfo{volume}{284} (\bibinfo{year}{2020}).
\bibitem[{Denner and {van Wachem}(2015)}]{Denner2015}
\bibinfo{author}{F.~Denner}, \bibinfo{author}{B.~{van Wachem}},
\newblock \bibinfo{title}{Numerical time-step restrictions as a result of
  capillary waves},
\newblock \bibinfo{journal}{Journal of Computational Physics}
  \bibinfo{volume}{285} (\bibinfo{year}{2015}) \bibinfo{pages}{24--40}.
\bibitem[{Popinet(2018)}]{Popinet2018}
\bibinfo{author}{S.~Popinet},
\newblock \bibinfo{title}{Numerical models of surface tension},
\newblock \bibinfo{journal}{Annual Review of Fluid Mechanics}
  \bibinfo{volume}{50} (\bibinfo{year}{2018}) \bibinfo{pages}{49--75}.
\bibitem[{Denner et~al.(2022)Denner, Evrard, and {van Wachem}}]{Denner2022b}
\bibinfo{author}{F.~Denner}, \bibinfo{author}{F.~Evrard},
  \bibinfo{author}{B.~{van Wachem}},
\newblock \bibinfo{title}{Breaching the capillary time-step constraint using a
  coupled {{VOF}} method with implicit surface tension},
\newblock \bibinfo{journal}{Journal of Computational Physics}
  \bibinfo{volume}{459} (\bibinfo{year}{2022}) \bibinfo{pages}{111128}.
\bibitem[{Janodet et~al.(2025)Janodet, {van Wachem}, and Denner}]{Janodet2025}
\bibinfo{author}{R.~Janodet}, \bibinfo{author}{B.~{van Wachem}},
  \bibinfo{author}{F.~Denner},
\newblock \bibinfo{title}{A fully-coupled algorithm with implicit surface
  tension treatment for interfacial flows with large density ratios},
\newblock \bibinfo{journal}{Journal of Computational Physics}
  \bibinfo{volume}{520} (\bibinfo{year}{2025}) \bibinfo{pages}{113520}.
\bibitem[{{Phan-Thien} and Tanner(1977)}]{Phan-Thien1977}
\bibinfo{author}{N.~{Phan-Thien}}, \bibinfo{author}{R.~I. Tanner},
\newblock \bibinfo{title}{A new constitutive equation derived from network
  theory},
\newblock \bibinfo{journal}{Journal of Non-Newtonian Fluid Mechanics}
  \bibinfo{volume}{2} (\bibinfo{year}{1977}) \bibinfo{pages}{353--365}.
\bibitem[{Giesekus(1982)}]{Giesekus1982}
\bibinfo{author}{H.~Giesekus},
\newblock \bibinfo{title}{A simple constitutive equation for polymer fluids
  based on the concept of deformation-dependent tensorial mobility},
\newblock \bibinfo{journal}{Journal of Non-Newtonian Fluid Mechanics}
  \bibinfo{volume}{11} (\bibinfo{year}{1982}) \bibinfo{pages}{69--109}.
\bibitem[{Gorges et~al.(2023)Gorges, Hod{\v z}i{\'c}, Evrard, {van Wachem},
  Velte, and Denner}]{Gorges2023}
\bibinfo{author}{C.~Gorges}, \bibinfo{author}{A.~Hod{\v z}i{\'c}},
  \bibinfo{author}{F.~Evrard}, \bibinfo{author}{B.~{van Wachem}},
  \bibinfo{author}{C.~M. Velte}, \bibinfo{author}{F.~Denner},
\newblock \bibinfo{title}{Efficient reduction of vertex clustering using front
  tracking with surface normal propagation restriction},
\newblock \bibinfo{journal}{Journal of Computational Physics}
  \bibinfo{volume}{491} (\bibinfo{year}{2023}) \bibinfo{pages}{112406}.
\bibitem[{Snoeijer et~al.(2020)Snoeijer, Pandey, Herrada, and
  Eggers}]{Snoeijer2020}
\bibinfo{author}{J.~H. Snoeijer}, \bibinfo{author}{A.~Pandey},
  \bibinfo{author}{M.~A. Herrada}, \bibinfo{author}{J.~Eggers},
\newblock \bibinfo{title}{The relationship between viscoelasticity and
  elasticity},
\newblock \bibinfo{journal}{Proceedings of the Royal Society A: Mathematical,
  Physical and Engineering Sciences} \bibinfo{volume}{476}
  (\bibinfo{year}{2020}) \bibinfo{pages}{20200419}.
\bibitem[{Alves et~al.(2001)Alves, Pinho, and Oliveira}]{Alves2001}
\bibinfo{author}{M.~A. Alves}, \bibinfo{author}{F.~T. Pinho},
  \bibinfo{author}{P.~J. Oliveira},
\newblock \bibinfo{title}{Study of steady pipe and channel flows of a
  single-mode {{Phan-Thien}}--{{Tanner}} fluid},
\newblock \bibinfo{journal}{Journal of Non-Newtonian Fluid Mechanics}
  \bibinfo{volume}{101} (\bibinfo{year}{2001}) \bibinfo{pages}{55--76}.
\bibitem[{Bartholomew et~al.(2018)Bartholomew, Denner, {Abdol-Azis}, Marquis,
  and {van Wachem}}]{Bartholomew2018}
\bibinfo{author}{P.~Bartholomew}, \bibinfo{author}{F.~Denner},
  \bibinfo{author}{M.~{Abdol-Azis}}, \bibinfo{author}{A.~Marquis},
  \bibinfo{author}{B.~{van Wachem}},
\newblock \bibinfo{title}{Unified formulation of the momentum-weighted
  interpolation for collocated variable arrangements},
\newblock \bibinfo{journal}{Journal of Computational Physics}
  \bibinfo{volume}{375} (\bibinfo{year}{2018}) \bibinfo{pages}{177--208}.
\bibitem[{Moukalled et~al.(2016)Moukalled, Mangani, and
  Darwish}]{Moukalled2016}
\bibinfo{author}{F.~Moukalled}, \bibinfo{author}{L.~Mangani},
  \bibinfo{author}{M.~Darwish}, \bibinfo{title}{The Finite Volume Method in
  Computational Fluid Dynamics: {{An}} Advanced Introduction with {{OpenFOAM}}
  and {{Matlab}}}, \bibinfo{publisher}{Springer}, \bibinfo{year}{2016}.
\bibitem[{Alves et~al.(2003)Alves, Oliveira, and Pinho}]{Alves2003}
\bibinfo{author}{M.~A. Alves}, \bibinfo{author}{P.~J. Oliveira},
  \bibinfo{author}{F.~T. Pinho},
\newblock \bibinfo{title}{A convergent and universally bounded interpolation
  scheme for the treatment of advection},
\newblock \bibinfo{journal}{International Journal for Numerical Methods in
  Fluids} \bibinfo{volume}{41} (\bibinfo{year}{2003}) \bibinfo{pages}{47--75}.
\bibitem[{Demird{\v z}i{\'c} and Muzaferija(1995)}]{Demirdzic1995}
\bibinfo{author}{I.~Demird{\v z}i{\'c}}, \bibinfo{author}{S.~Muzaferija},
\newblock \bibinfo{title}{Numerical method for coupled fluid flow, heat
  transfer and stress analysis using unstructured moving meshes with cells of
  arbitrary topology},
\newblock \bibinfo{journal}{Computer Methods in Applied Mechanics and
  Engineering} \bibinfo{volume}{125} (\bibinfo{year}{1995})
  \bibinfo{pages}{235--255}.
\bibitem[{Mathur and Murthy(1997)}]{Mathur1997}
\bibinfo{author}{S.~Mathur}, \bibinfo{author}{J.~Murthy},
\newblock \bibinfo{title}{A pressure-based method for unstructured meshes},
\newblock \bibinfo{journal}{Numerical Heat Transfer Part B Fundamentals}
  \bibinfo{volume}{31} (\bibinfo{year}{1997}) \bibinfo{pages}{195--215}.
\bibitem[{Ferziger(2003)}]{Ferziger2003}
\bibinfo{author}{J.~Ferziger},
\newblock \bibinfo{title}{Interfacial transfer in {{Tryggvason}}'s method},
\newblock \bibinfo{journal}{International Journal for Numerical Methods in
  Fluids} \bibinfo{volume}{41} (\bibinfo{year}{2003})
  \bibinfo{pages}{551--560}.
\bibitem[{Mencinger and Zun(2007)}]{Mencinger2007}
\bibinfo{author}{J.~Mencinger}, \bibinfo{author}{I.~Zun},
\newblock \bibinfo{title}{On the finite volume discretization of discontinuous
  body force field on collocated grid: {{Application}} to {{VOF}} method},
\newblock \bibinfo{journal}{Journal of Computational Physics}
  \bibinfo{volume}{221} (\bibinfo{year}{2007}) \bibinfo{pages}{524--538}.
\bibitem[{Gu{\'e}nette and Fortin(1995)}]{Guenette1995}
\bibinfo{author}{R.~Gu{\'e}nette}, \bibinfo{author}{M.~Fortin},
\newblock \bibinfo{title}{A new mixed finite element method for computing
  viscoelastic flows},
\newblock \bibinfo{journal}{Journal of Non-Newtonian Fluid Mechanics}
  \bibinfo{volume}{60} (\bibinfo{year}{1995}) \bibinfo{pages}{27--52}.
\bibitem[{Pimenta and Alves(2017)}]{Pimenta2017}
\bibinfo{author}{F.~Pimenta}, \bibinfo{author}{M.~Alves},
\newblock \bibinfo{title}{Stabilization of an open-source finite-volume solver
  for viscoelastic fluid flows},
\newblock \bibinfo{journal}{Journal of Non-Newtonian Fluid Mechanics}
  \bibinfo{volume}{239} (\bibinfo{year}{2017}) \bibinfo{pages}{85--104}.
\bibitem[{Rhie and Chow(1983)}]{Rhie1983}
\bibinfo{author}{C.~M. Rhie}, \bibinfo{author}{W.~L. Chow},
\newblock \bibinfo{title}{Numerical study of the turbulent flow past an airfoil
  with trailing edge separation},
\newblock \bibinfo{journal}{AIAA Journal} \bibinfo{volume}{21}
  (\bibinfo{year}{1983}) \bibinfo{pages}{1525--1532}.
\bibitem[{Peskin(1977)}]{Peskin1977}
\bibinfo{author}{C.~S. Peskin},
\newblock \bibinfo{title}{Numerical analysis of blood flow in the heart},
\newblock \bibinfo{journal}{Journal of Computational Physics}
  \bibinfo{volume}{25} (\bibinfo{year}{1977}) \bibinfo{pages}{220--252}.
\bibitem[{Tryggvason et~al.(2011)Tryggvason, Scardovelli, and
  Zaleski}]{Tryggvason2011a}
\bibinfo{author}{G.~Tryggvason}, \bibinfo{author}{R.~Scardovelli},
  \bibinfo{author}{S.~Zaleski}, \bibinfo{title}{Direct Numerical Simulations of
  Gas-Liquid Multiphase Flows}, \bibinfo{publisher}{Cambridge University
  Press}, \bibinfo{address}{Cambridge ; New York}, \bibinfo{year}{2011}.
\bibitem[{Pivello(2012)}]{Pivello2012}
\bibinfo{author}{M.~R. Pivello}, \bibinfo{title}{A Fully Adaptive
  Front-Tracking Method for the Simulation of {{3D}} Two-Phase Flows}, Ph.D.
  thesis, University of Uberlandia, \bibinfo{address}{Uberlandia},
  \bibinfo{year}{2012}.
\bibitem[{{de Sousa} et~al.(2004){de Sousa}, Mangiavacchi, Nonato, Castelo,
  Tom{\'e}, Ferreira, Cuminato, and McKee}]{deSousa2004}
\bibinfo{author}{F.~{de Sousa}}, \bibinfo{author}{N.~Mangiavacchi},
  \bibinfo{author}{L.~Nonato}, \bibinfo{author}{A.~Castelo},
  \bibinfo{author}{M.~Tom{\'e}}, \bibinfo{author}{V.~Ferreira},
  \bibinfo{author}{J.~Cuminato}, \bibinfo{author}{S.~McKee},
\newblock \bibinfo{title}{A front-tracking/front-capturing method for the
  simulation of {{3D}} multi-fluid flows with free surfaces},
\newblock \bibinfo{journal}{Journal of Computational Physics}
  \bibinfo{volume}{198} (\bibinfo{year}{2004}) \bibinfo{pages}{469--499}.
\bibitem[{Balay et~al.(2017{\natexlab{a}})Balay, Abhyankar, Adams, Brown,
  Brune, Buschelman, Dalcin, Eijkhout, Kaushik, Knepley, May, McInnes, Gropp,
  Rupp, Sanan, Smith, Zampini, Zhang, and Zhang}]{petsc-user-ref}
\bibinfo{author}{S.~Balay}, \bibinfo{author}{S.~Abhyankar},
  \bibinfo{author}{M.~F. Adams}, \bibinfo{author}{J.~Brown},
  \bibinfo{author}{P.~Brune}, \bibinfo{author}{K.~Buschelman},
  \bibinfo{author}{L.~Dalcin}, \bibinfo{author}{V.~Eijkhout},
  \bibinfo{author}{D.~Kaushik}, \bibinfo{author}{M.~G. Knepley},
  \bibinfo{author}{D.~A. May}, \bibinfo{author}{L.~C. McInnes},
  \bibinfo{author}{W.~D. Gropp}, \bibinfo{author}{K.~Rupp},
  \bibinfo{author}{P.~Sanan}, \bibinfo{author}{B.~F. Smith},
  \bibinfo{author}{S.~Zampini}, \bibinfo{author}{H.~Zhang},
  \bibinfo{author}{H.~Zhang}, \bibinfo{title}{{{PETSc Users Manual}}},
  \bibinfo{type}{Technical Report} \bibinfo{number}{ANL-95/11 - Revision 3.8},
  Argonne National Laboratory, \bibinfo{year}{2017}{\natexlab{a}}.
\bibitem[{Balay et~al.(2017{\natexlab{b}})Balay, Abhyankar, Adams, Brown,
  Brune, Buschelman, Dalcin, Eijkhout, Gropp, Kaushik, Knepley, McInnes, Rupp,
  Smith, Zampini, Zhang, and Zhang}]{petsc-web-page}
\bibinfo{author}{S.~Balay}, \bibinfo{author}{S.~Abhyankar},
  \bibinfo{author}{M.~F. Adams}, \bibinfo{author}{J.~Brown},
  \bibinfo{author}{P.~Brune}, \bibinfo{author}{K.~Buschelman},
  \bibinfo{author}{L.~Dalcin}, \bibinfo{author}{V.~Eijkhout},
  \bibinfo{author}{W.~D. Gropp}, \bibinfo{author}{D.~Kaushik},
  \bibinfo{author}{M.~G. Knepley}, \bibinfo{author}{L.~C. McInnes},
  \bibinfo{author}{K.~Rupp}, \bibinfo{author}{B.~F. Smith},
  \bibinfo{author}{S.~Zampini}, \bibinfo{author}{H.~Zhang},
  \bibinfo{author}{H.~Zhang}, \bibinfo{title}{{{PETSc Web}} page},
  \bibinfo{howpublished}{http://www.mcs.anl.gov/petsc},
  \bibinfo{year}{2017}{\natexlab{b}}.
\bibitem[{Yapici(2012)}]{Yapici2012}
\bibinfo{author}{K.~Yapici},
\newblock \bibinfo{title}{A comparison study on high-order bounded schemes:
  {{Flow}} of {{PTT-linear}} fluid in a lid-driven square cavity},
\newblock \bibinfo{journal}{Korea-Australia Rheology Journal}
  \bibinfo{volume}{24} (\bibinfo{year}{2012}) \bibinfo{pages}{11--21}.
\bibitem[{Ham and Iaccarino(2004)}]{Ham2004}
\bibinfo{author}{F.~Ham}, \bibinfo{author}{G.~Iaccarino},
\newblock \bibinfo{title}{Energy conservation in collocated discretization
  schemes on unstructured meshes},
\newblock \bibinfo{journal}{Annual Research Briefs, Center for Turbulence}
  (\bibinfo{year}{2004}) \bibinfo{pages}{3--14}.
\bibitem[{Pilz and Brenn(2007)}]{Pilz2007}
\bibinfo{author}{C.~Pilz}, \bibinfo{author}{G.~Brenn},
\newblock \bibinfo{title}{On the critical bubble volume at the rise velocity
  jump discontinuity in viscoelastic liquids},
\newblock \bibinfo{journal}{Journal of Non-Newtonian Fluid Mechanics}
  \bibinfo{volume}{145} (\bibinfo{year}{2007}) \bibinfo{pages}{124--138}.
\bibitem[{Bothe(2022)}]{Bothe2022}
\bibinfo{author}{D.~Bothe},
\newblock \bibinfo{title}{Sharp-interface continuum thermodynamics of
  multicomponent fluid systems with interfacial mass},
\newblock \bibinfo{journal}{International Journal of Engineering Science}
  \bibinfo{volume}{179} (\bibinfo{year}{2022}) \bibinfo{pages}{103731}.

\end{thebibliography}

\end{document}